\documentclass[reprint,superscriptaddress,nofootinbib,twocolumn,aps,amsmath,amssymb,pra,floatfix,10pt,eqsecnum]{revtex4-1}

\usepackage{hyperref}
\hypersetup{colorlinks=true, urlcolor=blue}

\usepackage[utf8]{inputenc}
\usepackage{bbm}
\usepackage{latexsym}
\usepackage{calc}
\usepackage{fancybox}
\usepackage{graphicx}  
\usepackage{subfigure}
\usepackage{verbatim}
\usepackage{todonotes}
\usepackage{amsthm} 

\newcommand{\RR}{\mathbb{R}}
\newcommand{\CC}{\mathbb{C}}
\newcommand{\NN}{\mathbb{N}}
\newcommand{\PP}{\mathbb{P}}
\newcommand{\EE}{\mathbb{E}}

\newcommand{\Cov}[2]{{\rm {Cov}} \left[ #1, #2 \right]}
\newcommand{\ket}[1]{\left | #1\right \rangle}
\newcommand{\bra}[1]{\left \langle #1 \right |}
\newcommand{\braket}[2]{\left \langle #1 | #2 \right \rangle}

\newcommand{\kphit}{\ket{\phi_t}}
\newcommand{\kpsit}{\ket{\psi_t}}

\newcommand{\dkpsit}{\ket{d\psi_t}}
\newcommand{\dkphit}{\ket{d\phi_t}}

\newcommand{\bphit}{\bra{\phi_t}}
\newcommand{\bpsit}{\bra{\psi_t}}

\newcommand{\Frt}[1]{F_{#1}^>(\theta_t)}
\newcommand{\Fr}[1]{F_{#1}^>(\theta)}
\newcommand{\Yrt}[1]{Y_{#1}^>(\theta_t)}
\newcommand{\Mrt}[1]{H_{#1}^>(\theta_t)}
\newcommand{\yrt}[1]{y_{#1}(\theta_t)}
\newcommand{\mrt}[1]{h_{#1}(\theta_t)}
\newcommand{\mrtinv}[1]{\left[h(\theta_t)^{-1}\right]^{#1}}
\newcommand{\Xre}[1]{X_{#1}^>(\eta)}
\newcommand{\Xle}[1]{X_{#1}^<(\eta)}

\newcommand{\CF}{\mathcal{J}}
\newcommand{\CGF}{\mathcal{J}^{\text{CGF}}}

\newcommand{\expect}[1]{\,\left\langle #1 \right\rangle}
\newcommand{\qcor}[2]{\,\sigma\left( #1, #2 \right)}
\newcommand{\cov}[2]{{\,\rm {cov}}\left( #1, #2 \right)}
\newcommand{\Var}[1]{{\,\rm {var}}\left( #1 \right)}

\newcommand{\partialD}[2]{\frac{\partial #1}{\partial #2}}

\newcommand{\re}[1]{{\,\rm {Re}} \left[#1\right]}
\newcommand{\im}[1]{{\,\rm {Im}} \left[#1\right]}

\newcommand{\Tr}[1]{{\rm Tr}\left(#1\right)}
\newcommand{\nr}{\nonumber}

\newcommand{\sigmatest}{U_\theta^\dagger \rho U_{\theta}}

\begin{document}

\title{Low dimensional manifolds for exact representation of open quantum systems}

\author{Nikolas Tezak}
\email{nikolas@rigetti.com}
\affiliation{Edward L. Ginzton Laboratory, Stanford University, Stanford, CA 94305, USA}
\affiliation{Rigetti Quantum Computing, Berkeley, CA 94702, USA}
\author{Nina H. Amini}
\affiliation{CNRS, Laboratoire des signaux et systèmes (L2S), CentraleSupélec, 3 rue Joliot Curie, 91192 Gif-Sur-Yvette, France}
\author{Hideo Mabuchi}
\affiliation{Edward L. Ginzton Laboratory, Stanford University, Stanford, CA 94305, USA}

\date{\today}

\begin{abstract}
Weakly nonlinear degrees of freedom in dissipative quantum systems tend to localize near manifolds of quasi-classical states. We present a family of analytical and computational methods for deriving optimal unitary model transformations that reduce the complexity of representing typical states. These transformations minimize the quantum relative entropy distance between a given state and particular quasi-classical manifolds. This naturally splits the description of quantum states into transformation coordinates that specify the nearest quasi-classical state and a transformed quantum state that can be represented in fewer basis levels. We derive coupled equations of motion for the coordinates and the transformed state and demonstrate how this can be exploited for efficient numerical simulation. Our optimization objective naturally quantifies the non-classicality of states occurring in some given open system dynamics. This allows us to compare the intrinsic complexity of different open quantum systems.
\end{abstract}

\maketitle


\section{Introduction} 
\label{sec:introduction}

A given quantum mechanical system can be described in more than one way. Our choice of description is usually motivated by the insight it provides, its economy, its accuracy and, when dealing with sufficient complexity, the efficiency with which it can be numerically simulated and analyzed.

Closed quantum systems evolve unitarily, and if their Hamilton operator admits a sufficient set of individually tunable control operators one can -- at least in principle -- realize arbitrary unitary operations on the system's Hilbert space \cite{Schirmer2001}. This enables powerful quantum computing and quantum simulation schemes that derive their advantage over classical computers from the exponential scaling of the Hilbert space dimension with system size.

We may then ask, what are the implications for open quantum systems which exhibit dissipative dynamics? As dissipation increases it becomes increasingly difficult to use them for unitary quantum computing, but there also exist applications in quantum engineering that explicitly require coupling to input and output fields, ranging from quantum limited signal amplification \cite{Caves1982Quantum,Mollow1967Quantum}, via quantum key distribution to autonomously correcting quantum memories \cite{Kerckhoff2010Designing}. An important class of such applications can be described in the language of quantum feedback networks \cite{Gough2008Quantum,Gough2009Series} or quantum input-output models \cite{Combes2016}. 

In general, our ability to \emph{design} quantum systems for specific tasks is severely limited by the state space complexity. This is true of closed and open systems, but as fewer guarantees exist on what dynamics are achievable with open quantum systems than in the closed system case we are even more reliant on efficient numerical schemes.
Fortunately, the dynamics of open quantum systems tend to exhibit phase space \emph{localization} \cite{Percival1999Localization}, which implies that there exist nonlinear sets within the system's Hilbert space that act as attractors for the quantum dynamics.

A very appealing feature of quantum network models based exclusively on nonlinear oscillators is that they allow continuously tuning between the coherent qubit regime \cite{Mabuchi2012Qubit} and the semi-classical weakly nonlinear limit. Several near-term applications exist for nonlinear oscillator networks: in the weakly nonlinear regime these include frameworks for photonic logic \cite{Mabuchi2011Nonlinear,Santori2014Quantum}, photonic Ising machines \cite{Wang2013Coherent,Inagaki2016LargeScale} and for all-optical machine learning \cite{Tezak2015Coherent}. In the strongly nonlinear regime novel quantum error correcting schemes have been proposed \cite{Mirrahimi2014Dynamically} and implemented \cite{Leghtas2015}. For all of these systems it is an extremeley interesting question to ask how their dynamics and capabilities change as the ratio of dissipation to non-linearity is increased or decreased, but no existing simulation method has allowed continuous interpolation between these regimes.
Our approach remediates this by providing a framework that relies on exact quantum model transformations to exploit semi-classical localization and obtain more efficient system representations. 

\begin{figure}[hbt]
    \centering
    \includegraphics[width=\columnwidth]{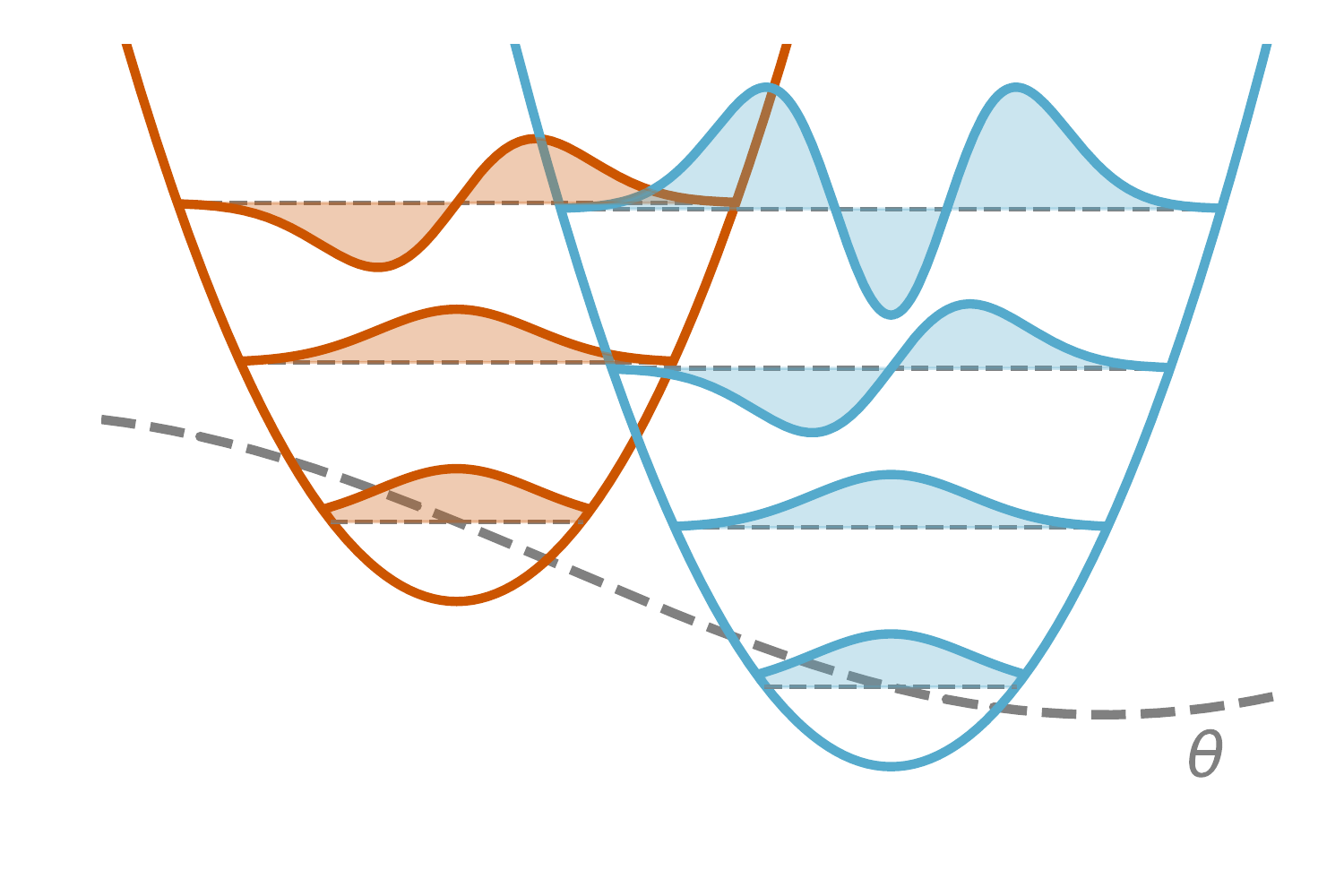}
    \caption{Smoothly parametrized unitary maps induce a family of localized orthonormal bases parametrized by the transformation coordinates. These correspond to semi-classical phase space variables. If the group representation is irreducible and the group semi-simple, then the `ground state' attached to each manifold point can be understood as a (generalized) coherent state \cite{Perelomov1977Generalized}.
    For open, diffusive quantum dynamics, quantum states often localize in the vicinity of such generalized coherent states.
    }
    \label{fig:qmanifold_visualization}
\end{figure}

The key to our method is formulating the problem of finding efficient state parametrizations as an optimization problem.
Specifically, we employ smoothly parametrized unitary transformations to represent states in co-moving adaptive bases. The description of, e.g., a pure quantum state $\kpsit$ thus splits into the transformation coordinates $\theta_t$ and a residual quantum state $\kphit = U_{\theta_t}^\dagger \kpsit$ that is represented in a localized basis. The diagram in Figure~\ref{fig:qmanifold_visualization} intends to visualize this. 
We quantify localization in terms of complexity functionals that are shown to have precise information geometric meaning. 

With this, we establish a self-consistent analytic framework that not only allows for numerical simulation but also serves to derive coupled dynamical equations for the semi-classical group manifold coordinates and the residual quantum state. We provide expressions for different diffusive Schrödinger equations as well as (stochastic) master equations. Although \emph{stochasticity} is not at all a necessary requirement of our approach, we expect our method to have the greatest impact for systems that are continuously weakly observed by the environment and this naturally leads us to consider stochastic dynamics. 

There already exist methods that rely on localization to analytically derive semi-classical stochastic dynamics \cite{Carter1995Quantum} from quantum quasi-probability distributions \cite{Wigner1932On,Husimi1940Some,Sudarshan1963Equivalence}. These, however, require approximations that can lead to significant discrepancies with simulations based on a full quantum state description \cite{Kinsler1991Quantum,Santori2014Quantum}. Furthermore, there does not exist a general approach to incrementally increase the accuracy of these methods.

Even prior to Percival's analytic results on localization, Schack, Brun and Percival developed the MQSD simulation method \cite{Schack1995Quantum,Steimle1999Mixed} which allowed simulating stochastic Schrödinger equations in displaced Fock bases. It can be seen as a special case of our approach, but it does not produce analytical equations of motion for the transformation coordinates. The MQSD simulation method works well for strongly dissipative oscillator systems that exhibit nearly Gaussian states with low amounts of squeezing or multi-mode correlations, but it does not generalize to other degrees of freedom. Furthermore, MQSD is based on a heuristic that has lacked a rigorous interpretation prior to our framework.

There have also been several \emph{model-agnostic} attempts to reduce the computational complexity of simulating open quantum systems: in \cite{Tezak2012Specification} a photonic SR-latch in the weakly quantum regime was approximated by a finite state-space Markov chain that could be re-expressed as an effective quantum network model, and in \cite{Shi2016Model} a conceptually very appealing approach was proposed based on principal component analysis of simulated quantum states. One important shortcoming of both of these methods, however, is that they require at least one simulation of the full system model which can be very expensive and in many cases impossible. Furthermore, they do not preserve important qualitative features of a quantum model, such as the commutators of system observables.

The contents of this paper are as follows:
First, we motivate the technical details presented in subsequent sections by presenting numerical results of our method applied to two separate systems: (a) the Kerr-nonlinearity based SR-latch as first described in \cite{Mabuchi2011Nonlinear} and discussed as an example in \cite{Tezak2012Specification} and \cite{Shi2016Model}, and (b) the degenerate parametric oscillator (DPO) model which is a highly interesting system both in the semi-classical regime where it exhibits a pitchfork bifurcation and in the strongly nonlinear quantum regime, where the semi-classical fix-points limit to low decoherence cat-states that can be used to encode a qubit \cite{Mirrahimi2014Dynamically}.

Next, we formally introduce the underlying `state compression' framework which relies on positive `penalty'-operators to quantify the complexity of quantum states. We show that their spectra can be directly related to achievable bounds on numerical truncation errors, and how this can be exploited for adapting our approach to situations in which Gaussian states poorly approximate the quantum dynamics.

We then describe how our method can be understood from an information theoretic point of view by relating our complexity functional to the quantum relative entropy. This allows to characterize the attractors near which the dynamics of a given system localize in terms of generalized Gibbs states generated by our specific choice of penalty operator which is smoothly transformed by our group.

We further present a small library of possible groups and penalty operators and finish by providing an outline of how this method can be extended to higher-dimensional Lie groups for which it is infeasible to derive the transformation differential in closed form.

There are many exciting future directions and valuable applications of our research, some of which we mention in this paper's conclusion.


\section{Applications} 
\label{sub:applications}
We first present some numerical results of our method. The technical details will be described in the following sections.
Our results were obtained using a custom software package QMANIFOLD \cite{Tezak2016QMANIFOLD} that not only facilitates analytic derivations but also allows to directly carry out numerical simulations. It uses QuTiP's \cite{Johansson2012Qutip} datastructures and interfaces with our existing package QNET \cite{Tezak2012Qnet} to automate various tedious symbolic calculations related to deriving the adjoint group representation and thus the derivation of the right generators $\Frt{j}$. 
Given a particular transformation $U_\theta$ and complexity functional (cf.~Section~\ref{sec:model_reduction_vs_model_compression}) it can compute the minimum complexity state $\kphit$ and coordinates $\theta_t$ for any input state $\kpsit.$ Furthermore, given a dynamical open system model parametrized by its internal Hamiltonian $H$ and some dissipation operators $\mathbf{L}$ it can directly derive and simulate stochastic complex quantum diffusion dynamics.

\subsection{Toy Model: An empty cavity} 
\label{sub:an_empty_cavity}
Consider a simple open cavity model described by a Hamiltonian $H=\hbar \omega a^\dagger a$ and a single dissipation operator $L=\sqrt{\kappa}a.$
Using the group of coherent displacements and linear excitation minimization (cf.~Section~\ref{sub:coherent_displacement}) of the canonical counting operator $N=a^\dagger a$ we apply the method of gradient coupled fiducial state dynamics with the fiducial state given by a displaced Gibbs state
\begin{align}
\chi_{\beta,\theta} = U_\theta \underbrace{Z(\beta)^{-1}e^{-\beta N}}_{\chi_\beta} U_\theta^\dagger
\end{align}
 with moving basis excitation $\expect{N}_{\chi_\beta} = \overline n_{\rm th}$. Summarizing the manifold coordinates in a single complex amplitude $\alpha = \frac{Q+iP}{\sqrt{2}}$, this coordinate and the reduced complexity state $\kphit$ have the joint equations of motion (cf Section~\ref{sec:dynamics_in_a_moving_basis})
\begin{align} \label{eq:dalpha_empty}
    d\alpha & = -\left[i\omega + \kappa/2\right]\alpha dt + \eta \expect{a}_\phi dt  + \sqrt{\kappa}\overline n_{\rm th}dW  \\
    \dkphit 
    \label{eq:dphi_empty} 
    & =  -  \left(i\omega +\frac{\kappa}{2}\right)  a^\dag a \kphit dt \\ \nr 
    & \quad -\eta \expect{a}_\phi a^\dagger \kphit dt + \eta \expect{a}_\phi^\ast a \kphit dt \\ \nr
    & \quad -  \sqrt{\kappa}\overline n_{\rm th}  a^\dagger \kphit dW
    +   \sqrt{\kappa} (\overline n_{\rm th} + 1) a\kphit dW^\ast 
\end{align}
where $\dkphit$ does not necessarily conserve the norm of $\kphit$ and where we have dropped a term $\propto \kphit dt$ that does not affect the evolution of any expectation values.

These dynamics can be simulated for any choice of $\overline n _{\rm th}$ which can also be turned into a dynamic variable (cf.~Section~\ref{sub:gibbs_manifold_projection}). For any finite choice of $\eta > 0$ the mode operator expectation will fluctuate around zero $\expect{a}_{\phi_t} \approx 0$ and in the limit $\eta \to \infty$ we recover the hard constraint $\left.\expect{a}_{\phi_t}\right|_{\eta \to \infty} = 0$.
We see that $\alpha$ only couples to the input noise process if the moving basis excitation is non-zero.
On the other hand, for $n_{\rm th}=0 \Leftrightarrow \beta \to \infty$ we find that $\kphit =\ket{0}$ is a stable fix-point of the dynamics in accordance with our intuition about passive linear quantum systems.
We could have equally well derived a (stochastic) master equation or a homodyne SSE using the formulas from Section \ref{sec:dynamics_in_a_moving_basis}.

We can easily add a coherent displacement to this model by modifying $L\to L +\epsilon, H\to H + \frac{\sqrt{\kappa}}{2i}(\epsilon a^\dagger - \epsilon^\ast a)$. A straightforward calculation reveals that this linear displacement is fully absorbed into the dynamics of $\alpha$ such that relative to Equations \eqref{eq:dalpha_empty},\eqref{eq:dphi_empty} we have
\begin{align}
    d\alpha \to d\alpha - \sqrt{\kappa} \epsilon dt
\end{align}
while $\dkphit$ is left unchanged.

\subsection{Kerr-cavity based NAND latch} 
\label{sub:kerr_cavity_based_nand_latch}
\begin{figure}[htb]
    \centering
    \includegraphics[width=0.8\columnwidth]{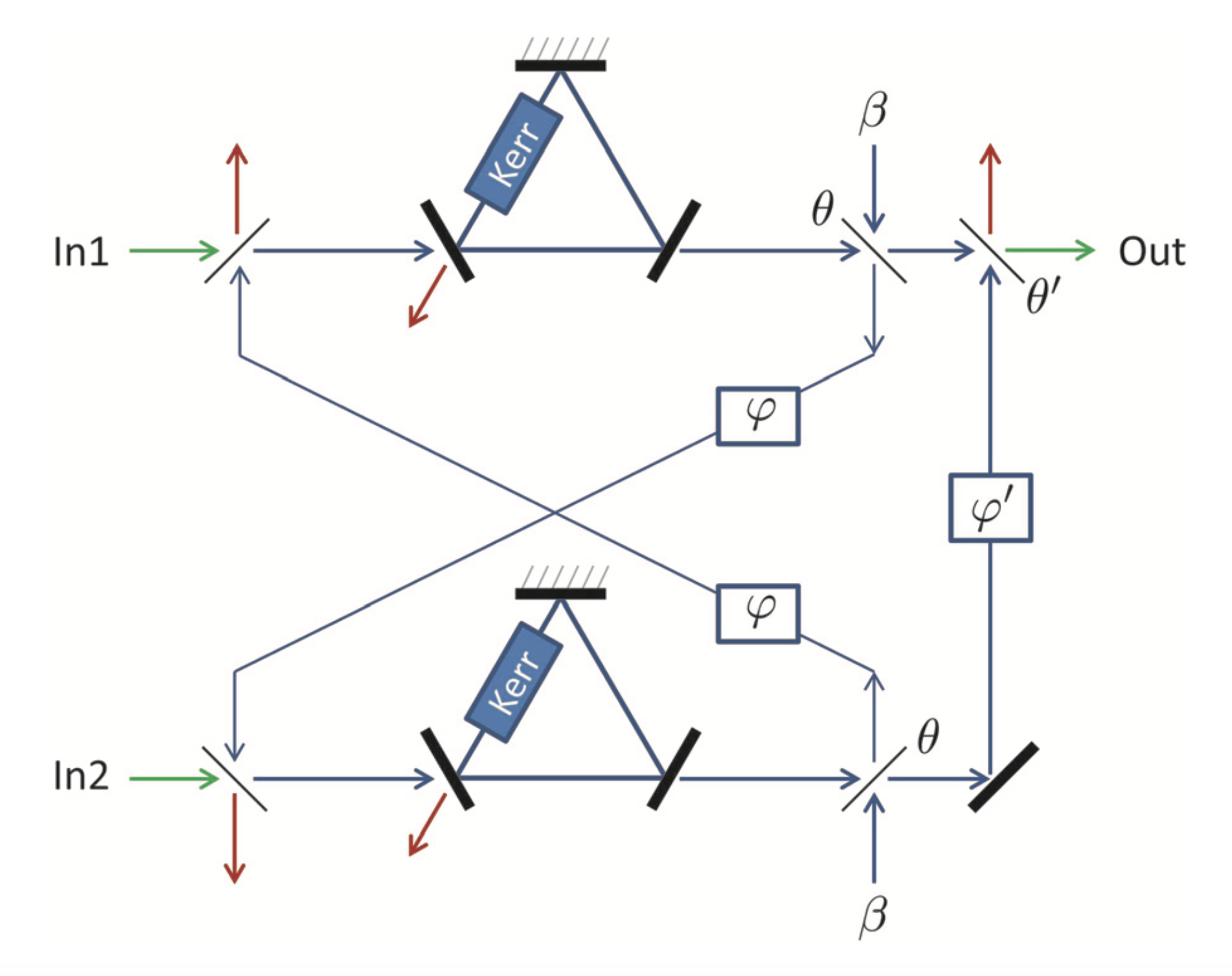}
    \caption{A coherent NAND latch as described in \cite{Mabuchi2011Nonlinear}.}
    \label{fig:nand_latch}
\end{figure}

 Mabuchi recently proposed \cite{Mabuchi2011Nonlinear} designing photonic logic gates by using nonlinear resonators in an interferometric feedback configuration. To achieve maximum power efficiency, devices could operate in a semi-classical regime where several tens of photons in a Kerr-resonator would cause the frequency to shift by one linewidth. A particularly interesting model is that of a photonic NAND-latch \cite{Mabuchi2011Nonlinear} symmetrically constructed from two Kerr-resonators with mutual coherent feedback as visualized in Figure~\ref{fig:nand_latch}. The circuit is designed such that it has two metastable states with either the first \emph{or} the second resonator in a high photon number state while the other resonator has low internal photon number.
A full quantum trajectory simulation of a Stochastic Schrödinger equation with $D= d^2 = 75^2=5625$ levels for this model with the same parameters as in \cite{Mabuchi2011Nonlinear} and \cite{Tezak2012Specification} is feasible on a typical workstation. 
\begin{figure*}[bt]
    \centering
    \includegraphics[width=\textwidth]{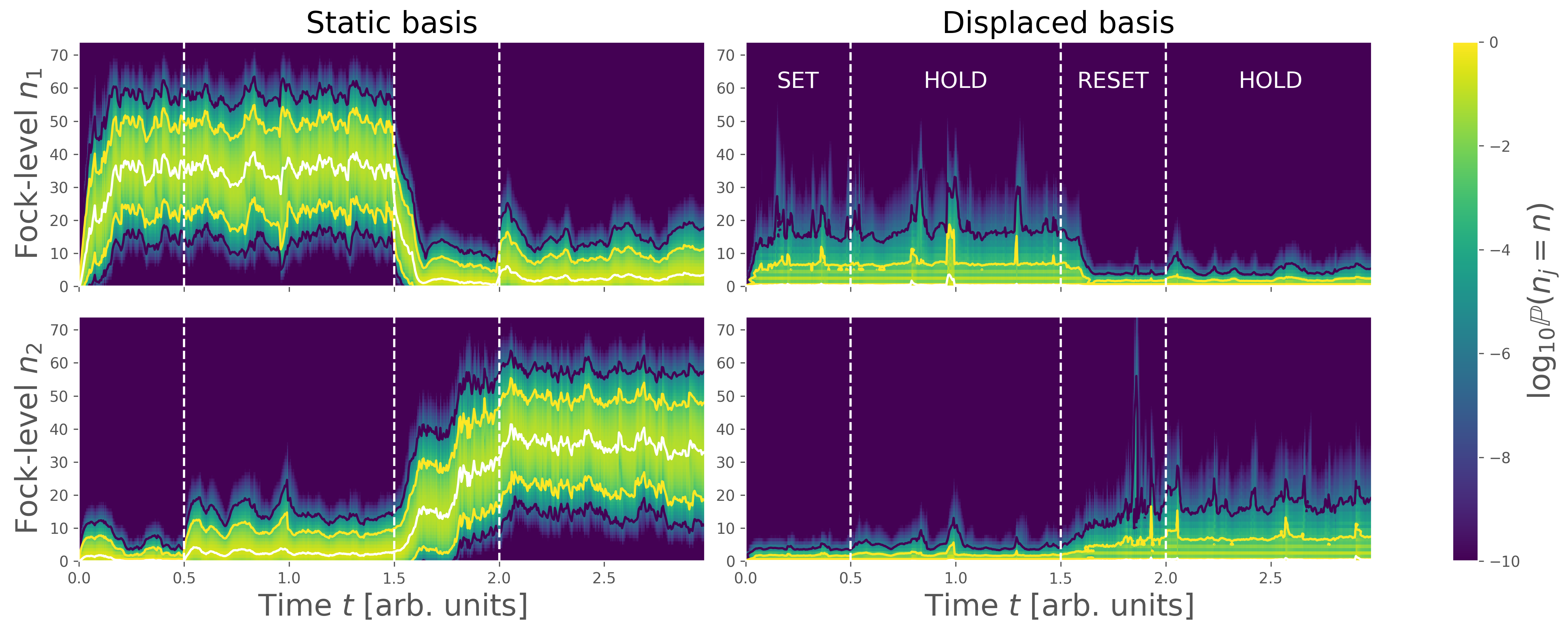}
    \caption{Heat map of the base-10 logarithm of the occupation probability of each oscillator's Fock levels. The left panels show the results in the static basis for the first oscillator (upper left) and the second oscillator (lower left). On the right hand side we present the results obtained in a jointly displaced basis for the first oscillator (upper right) and second oscillator (lower right). The overlaid white trace gives the expected number of excitations $\expect{a_j^\dagger a_j}$ in each basis. The yellow (dark blue) traces show the contours at which the occupation probability reaches $10^{-3}$ ($10^{-6}$).}
    \label{fig:nand_latch_reduction}
\end{figure*}
In Figure \ref{fig:nand_latch_reduction} we present the Fock level occupation probability $\mathbb{P}(n_j=n) := \expect{\Pi^{(j)}_n}$ for each oscillator mode $j=1,2$ and each Fock-level $n=0,1,\dots, d-1$ where $\Pi^{(j)}_n$ is the projection operator onto the $n$-th Fock state of oscillator $j$. We visualize this for both the original, static basis and for a coherently displaced representation, obtained by solving the linear excitation minimization problem for $N = a_1^\dagger a_1 + a_2^\dagger a_2$, i.e., the total number of excitations present in the displaced frame. 
We find that we can represent the dynamics of the system in a basis set that is between at least four and ten times smaller depending on whether we dynamically update the basis size during simulation \emph{without sacrificing any precision in the simulation.} If we further allow for modest simulation error, we can achieve up to an additional order of magnitude in basis reduction. Note that dimensional reduction for the same model was studied in \cite{Tezak2012Specification,Shi2016Model} and very similar physical models were studied in \cite{Santori2014Quantum} but we claim that ours is the only approach that allows to achieve strong dimensional reduction with a controlled error and without having to simulate the full system first.
\subsection{The degenerate parametric oscillator} 
\label{sub:example_application_the_degenerate_parametric_oscillator}

\begin{figure}[htb]
    \centering
    \includegraphics[width=0.95\columnwidth]{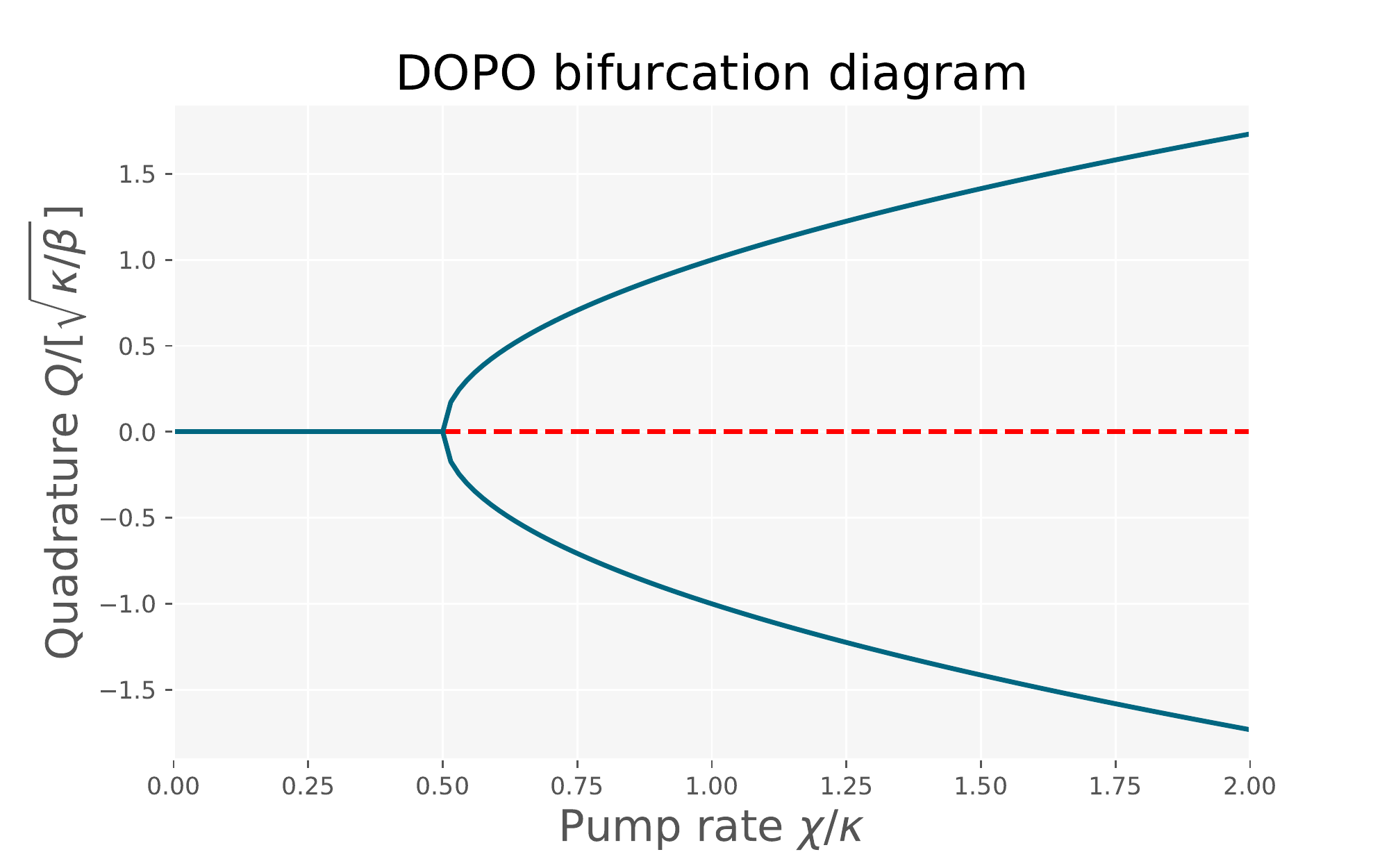}
    \caption{When the linear loss is larger than the gain $\kappa/2 < \chi$ the real quadrature $Q$ remains stably at $0$, above a critical pump $\chi \ge \kappa/2$ this fixpoint bifurcates into two stable symmetric solutions (solid lines) and an unstable solution (dashed line) that is the continuation of the below threshold $Q=0$ solution.}
    \label{fig:dopo_bifurc}
\end{figure}

\begin{figure}[htb]
    \centering
    \subfigure[$\beta/\kappa=1/12$]{\includegraphics[width=0.8\columnwidth]{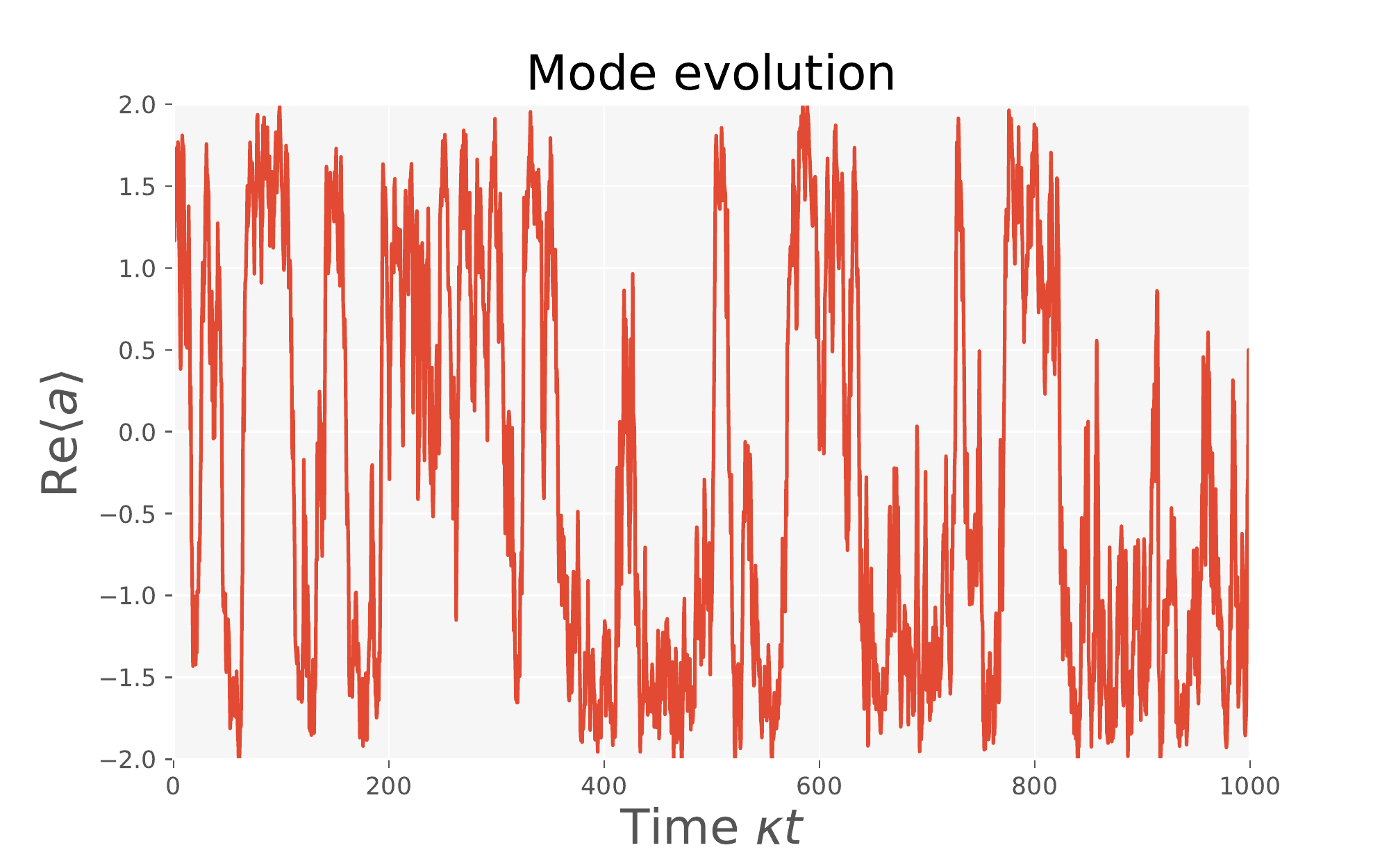}}
    \subfigure[$\beta/\kappa=1$]{\includegraphics[width=0.8\columnwidth]{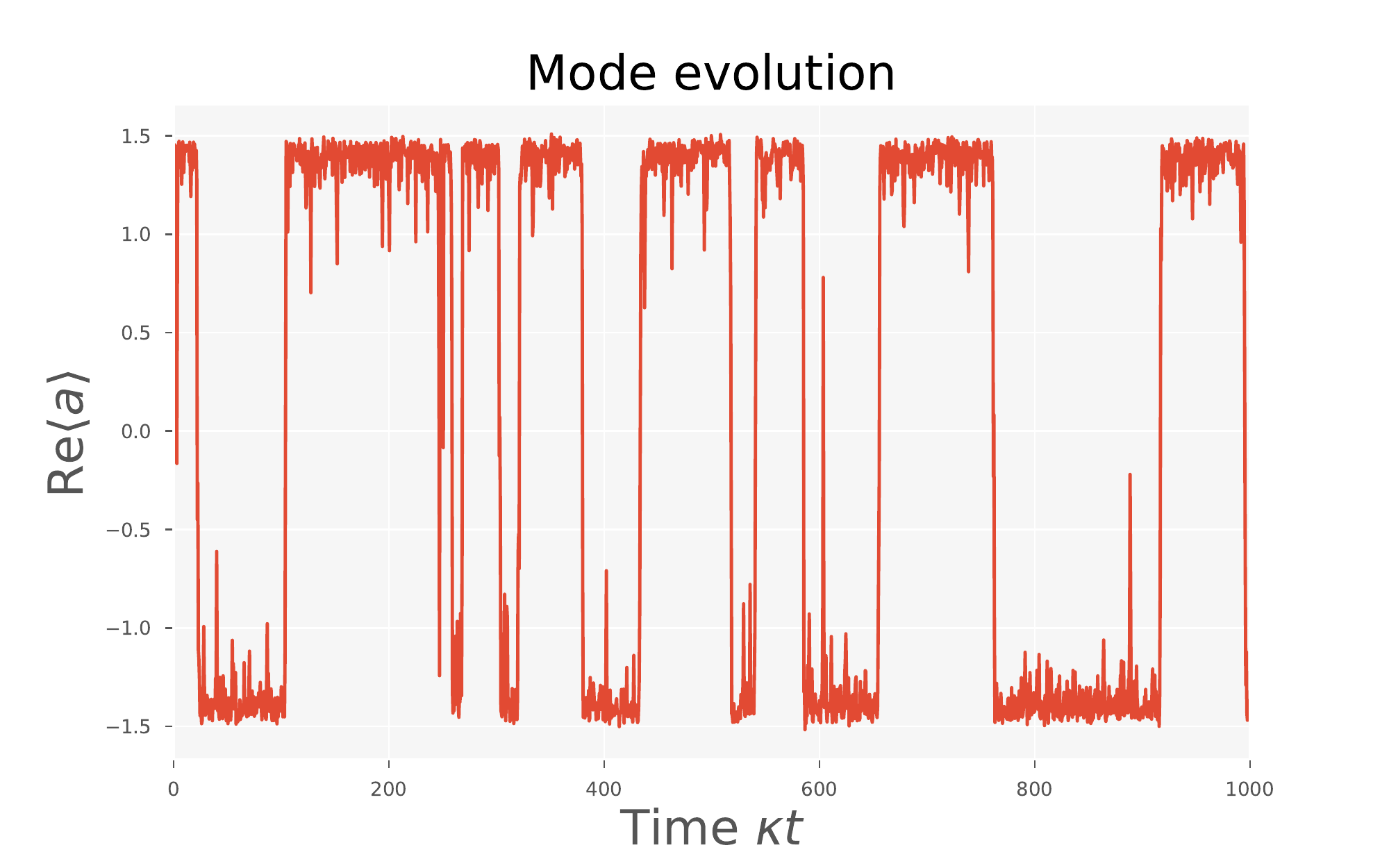}}
    \caption{Stochastic switching dynamics of a DPO above threshold. Figure (a) shows an example with very weak nonlinear loss $\beta \ll \kappa$ whereas (b) shows the strongly nonlinear case $\beta=\kappa.$
    In both cases we have chosen the parameters $\beta,\kappa,\chi$ such that the bi-stable mode amplitude equals approximately $\alpha_{r, ss}=\pm1/\sqrt{2}$. There is a visible reduction in the switching rate and we can also see quite clearly that the magnitude of fluctuations in either bi-stable state is strongly reduced in the case of very strong nonlinearity.
    Specifically, the simulation parameters were $\beta=\kappa,\chi=5\kappa/2$ and $\beta=\kappa/12,\chi=2\kappa/3$ for the strongly and weakly nonlinear case, respectively.
    }
    \label{fig:opo_switching}
\end{figure}

Another simple nonlinear extension of the empty cavity is given by the degenerate parametric oscillator (DPO). DPOs can exhibit very rich dynamics and have been long been employed for amplification \cite{akhmanov1965observation} and generation of light at tunable wavelengths \cite{Giordmaine1965Tunable} as well as more recently coherent optical Ising machines \cite{Utsunomiya2011Mapping}. It can be physically realized by a resonant signal mode coupled to a strongly driven pump mode through a nonlinear parametric interaction that mediates the conversion of signal photon pairs to pump photons and vice versa. In the strongly nonlinear limit and for a low quality factor of the pump mode, the pump mode can be adiabatically eliminated yielding an open system model in a rotating frame parametrized by
\begin{align}
    H &= i\frac{\hbar \chi}{2}\left[a^{\dagger 2}-a^2\right] \\
    \mathbf{L} &= \begin{pmatrix} \sqrt{\kappa}a \\ \sqrt{\beta} a^2 \end{pmatrix}.
\end{align}
Using the same Ansatz as above but with $\overline{n}_{\rm th} = \eta =0$ for simplicity we find for the deterministic part of the coordinate dynamics
\begin{align} \label{eq:dalpha_opo}
    d\alpha_{\rm det} & = -\left[\kappa/2+ \beta |\alpha|^2\right]\alpha dt + \chi \alpha^\ast dt.
\end{align}
For positive pump phase $\chi>0$ the dynamics are primarily captured by the evolution of the signal mode's real quadrature $Q = (\alpha+\alpha^\ast)/\sqrt{2}$ while the orthogonal $P$ quadrature is suppressed. We can then fix $P=0$ and consider the one dimensional differential equation valid in the classical limit
\begin{align}\label{eq:dopo_q_sc}
    \dot{Q} \approx -\left( \frac{\kappa}{2} - \chi + \frac{\beta}{2} Q^2 \right) Q.
\end{align}
Equation \eqref{eq:dopo_q_sc} is identical to the normal form of a pitchfork bifurcation up to some re-scaling. We visualize the bifurcation diagram in Figure~\ref{fig:dopo_bifurc}.
The bifurcation exists for any non-zero two-photon loss rate $\beta > 0,$ however the magnitude of $\beta$ strongly affects how non-classical (which in this context we take to mean non-Gaussian) the state of the signal mode becomes. When the system is pumped only slightly above the threshold, random switching or tunneling between the two equilibria is possible. We present such a trajectory in Figure \ref{fig:opo_switching}.
For a constant steady state mode amplitude the switching rate strongly depends on the ratio of linear to two-photon loss. Additionally, in the strongly nonlinear case $\beta \geq \kappa,$ the system can spontaneously evolve into cat-like states that feature exhibit significant simultaneous overlap with coherent states centered at either equilibrium.

In the limit of vanishing \emph{linear} loss $\kappa/\beta \to 0,$ the system has a decoherence-free sub-manifold spanned by the two equilibrium amplitude coherent states. In \cite{Mirrahimi2014Dynamically} Mirrahimi et al.~outline a scheme to encode quantum information in such a system.
A detailed study of the switching dynamics was carried out in \cite{Kinsler1991Quantum}.

\begin{figure}[htb]
    \centering
    \subfigure[$\beta/\kappa=1/12$]{\includegraphics[width=\columnwidth]{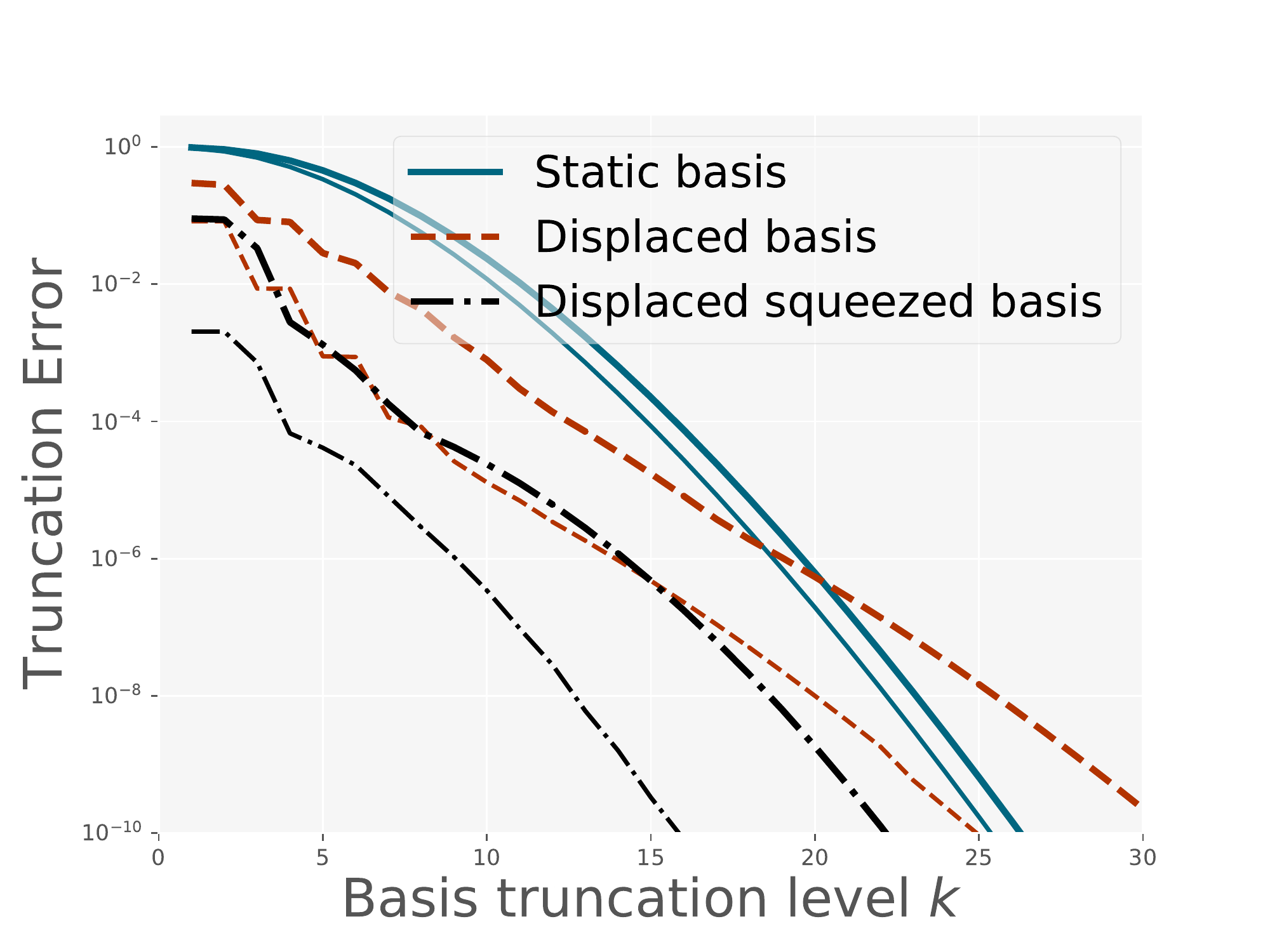}}
    \subfigure[$\beta/\kappa=1$]{\includegraphics[width=\columnwidth]{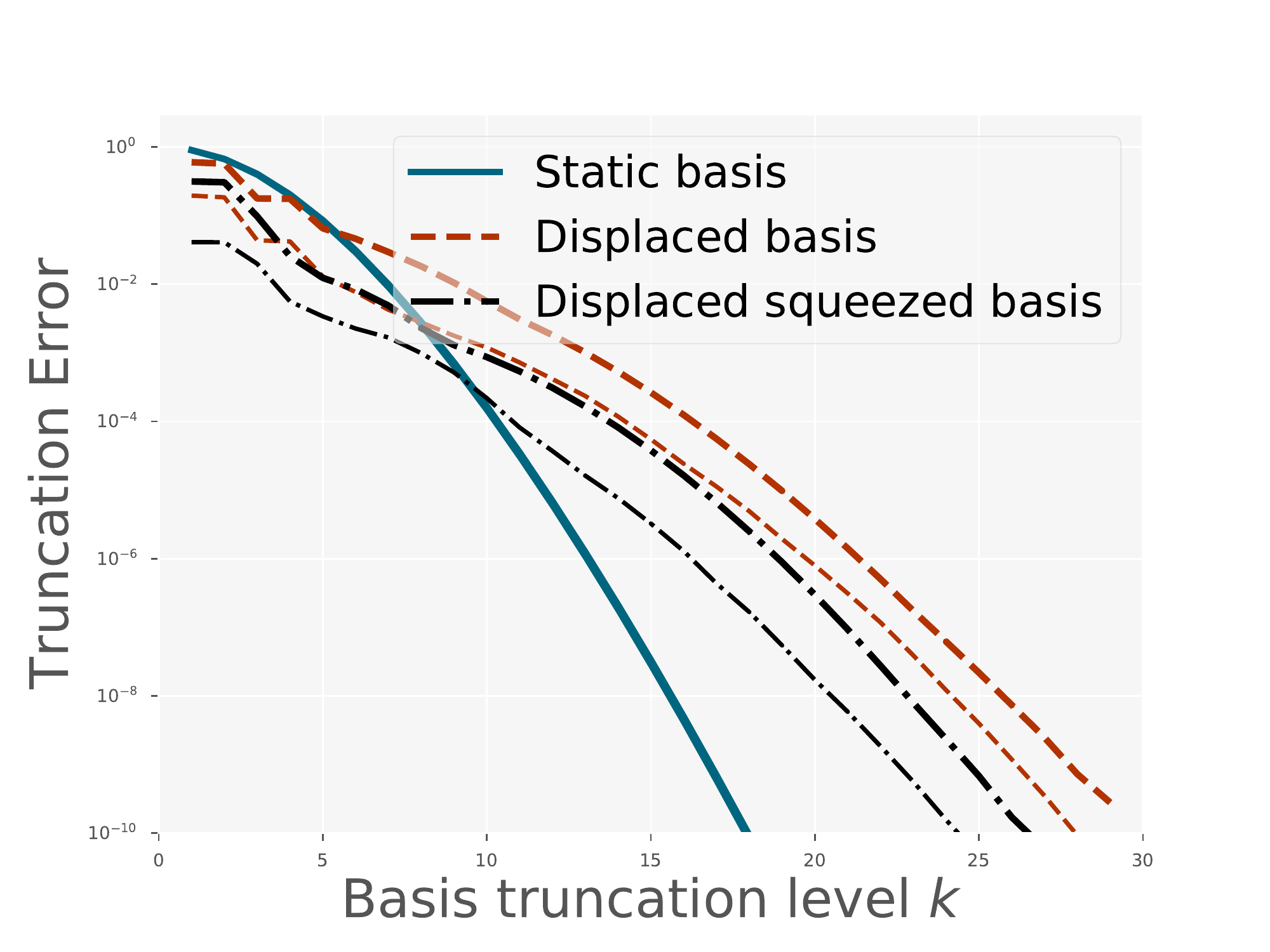}}
    \caption{For the weakly nonlinear case (a) both the displaced basis (dashed lines) and the displaced and squeezed basis (dash-dotted lines) perform fairly well, although the displaced basis truncation error falls off less rapidly than either the static or the displaced, squeezed basis. For strongly nonlinear case (b), however, we find that the static basis (solid lines) outperforms both the displaced (dashed) and the displaced, squeezed (dash-dotted) basis. This indicates that the system dynamics depart significantly from the squeezed and displaced coherent state manifold. In both figures the thinner lines of otherwise equal style indicate the 90\% level of the error, i.e., 90\% of all states had lower truncation error than that.}
    \label{fig:dopo_excitation_min}
\end{figure}

\begin{figure}[htb]
    \centering
   \subfigure[$\beta/\kappa=1/12$]{\includegraphics[width=\columnwidth]{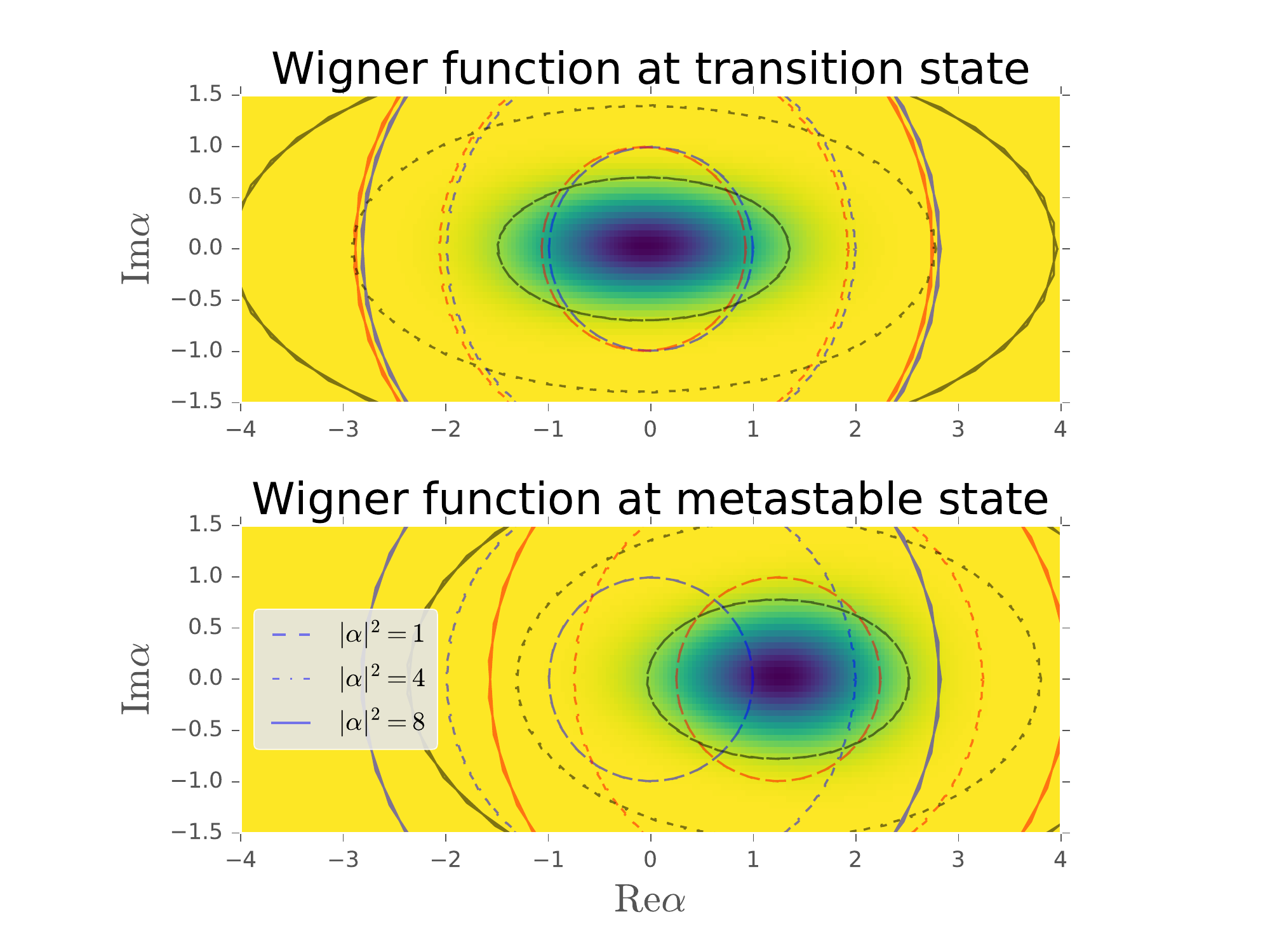}}
    \subfigure[$\beta/\kappa=1$]{\includegraphics[width=\columnwidth]{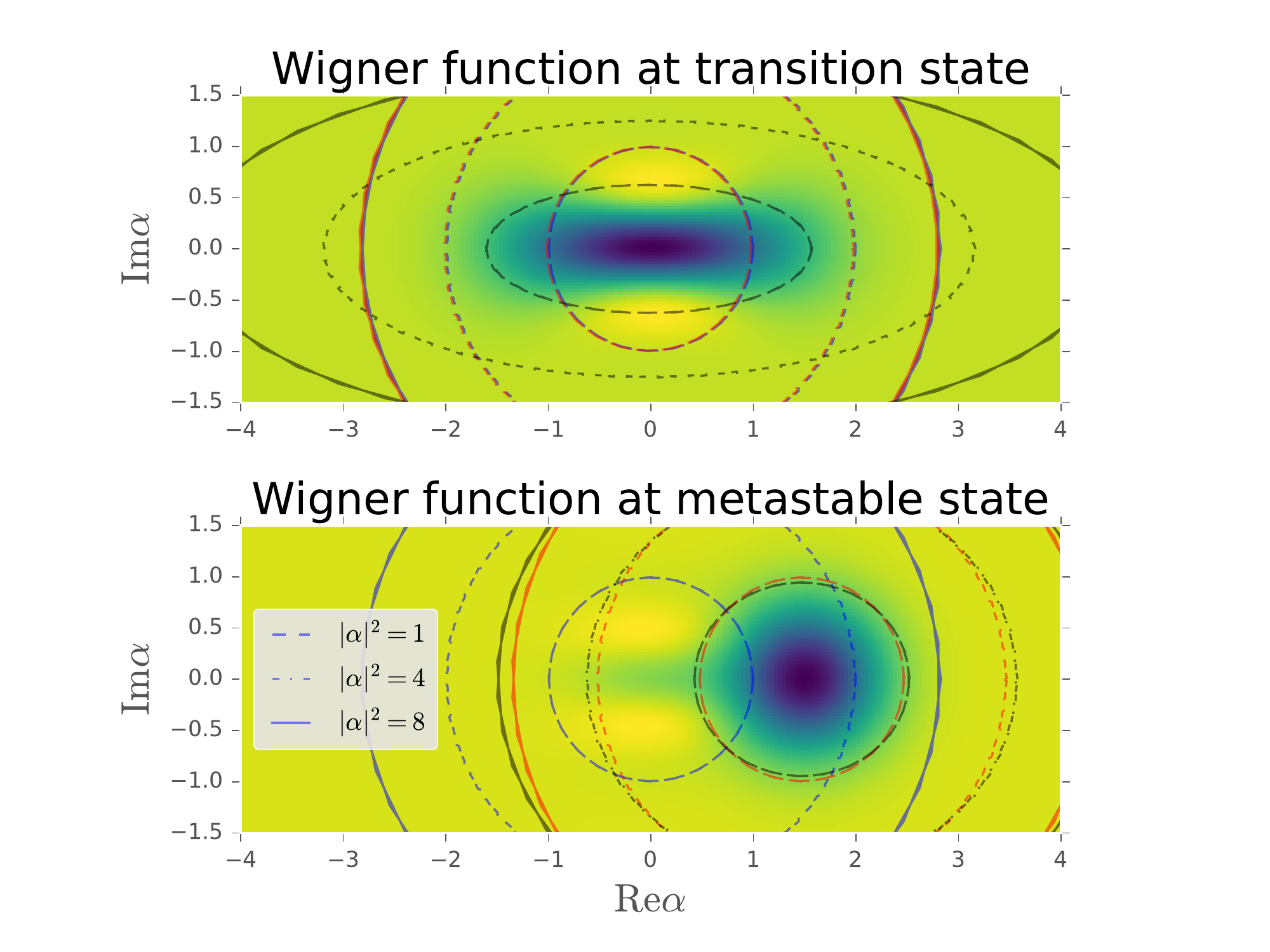}}
    \caption{Comparing the Wigner functions of either system in typical transitions states and typical meta-stable states we see clearly that the Wigner functions of the strongly nonlinear system (b) appear much less Gaussian in shape than for the system dominated by linear dissipation.
    We have furthermore indicated the support set of different bases. The blue circles correspond to the fixed basis, the red circles to a coherently displaced basis and the black ellipses to a displaced and squeezed basis.
    }
    \label{fig:dopo_excitation_min_wigner}
\end{figure}

\begin{figure}[htb]
    \centering
    \includegraphics[width=\columnwidth]{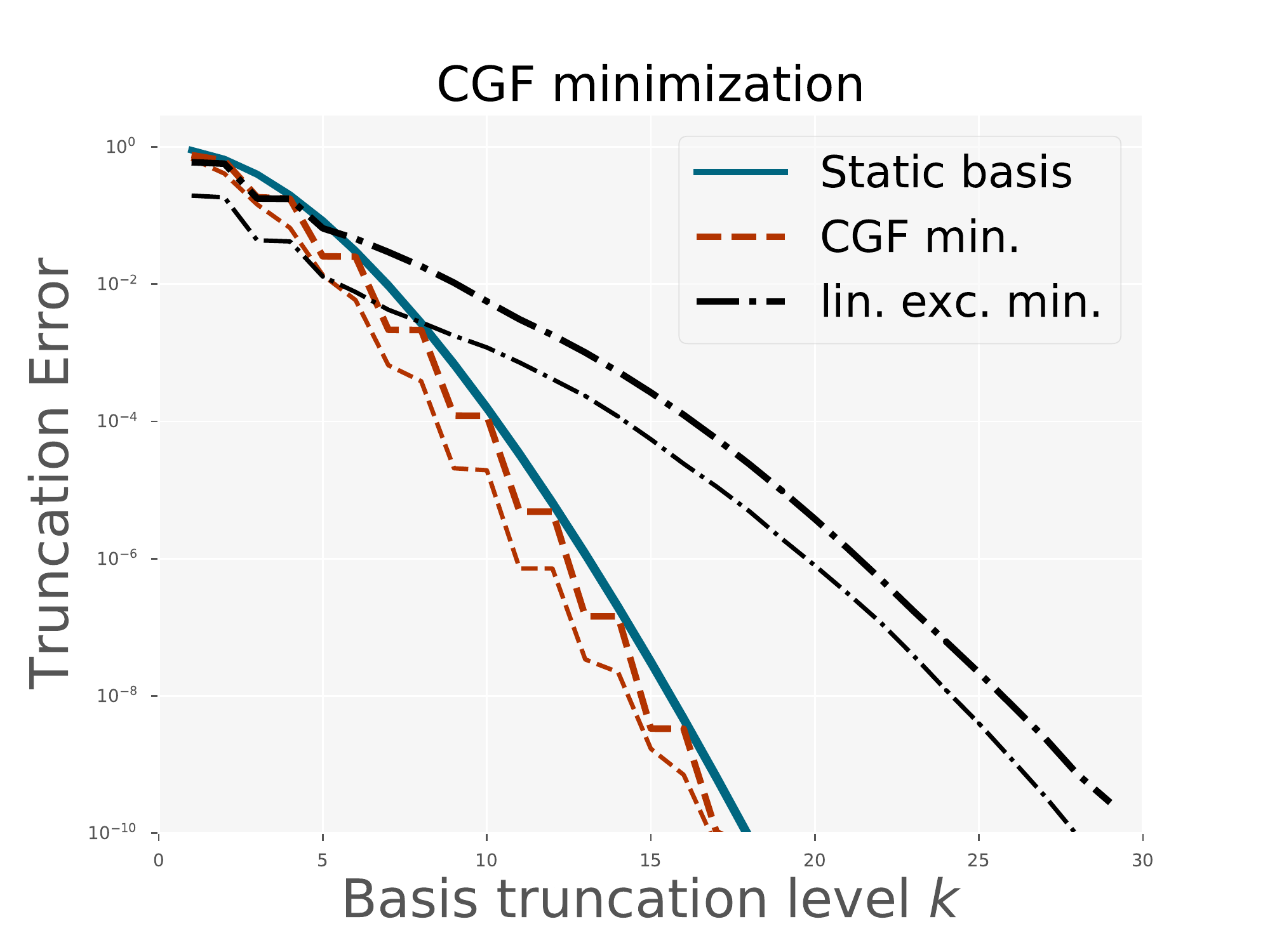}
    \caption{When changing the optimization problem to CGF minimization (dashed lines), we see that achieve higher representation efficiency than the static basis (solid lines) while so far only using a displaced, non-squeezed basis (dash-dotted lines). Again, the thinner lines of equal style indicate the 90\% levels.}
    \label{fig:dopo_CGF_min}
\end{figure}

In Figure \ref{fig:dopo_excitation_min} we compare how each basis level contributes to a whole trajectory of states when represented in the original fixed basis to one obtained via linear excitation minimization (cf.~Section~\ref{ssub:linear_excitation_minimization}) using either a coherently displaced basis or a displaced and squeezed basis. We see that the adaptive schemes perform well in the case of strong linear dissipation but not so well in the case of strong two-photon loss.
We can understand this better by inspecting typical states that occur in each evolution. In Figure \ref{fig:dopo_excitation_min_wigner} we present snapshots of the signal mode's Wigner function. For strong linear dissipation, the Wigner function of the signal mode typically appears quite Gaussian in shape, whereas in the strong two-photon loss case we see significant non-Gaussian features both in the transition states and when the mode is at one of the equilibria.
The bad performance of the excitation minimization functional in the non-Gaussian case is much improved by the cumulant generating function (CGF) minimization approach (cf.~Section~\ref{ssub:CGF_min}) . In Figure~\ref{fig:dopo_CGF_min} we compare the efficiency of the fixed basis with a coherently displaced basis where the coordinates are determined either by excitation minimization or by CGF minimization. We find that the CGF minimization (for $\lambda =3/2$) outperforms both the fixed basis and the excitation minimization method (which is equivalent to the QSD package's approach). Here we have not even exploited the additional advantages that a displaced \emph{and squeezed} basis combined with the CGF approach may yield.
Further details of the CGF minimized state representation can be found in Appendix \ref{sec:further_results_on_the_dopo_under_cgf_minimization}.

\section{Quantum state compression} 
\label{sec:model_reduction_vs_model_compression}

In this section we discuss different options for quantifying the efficiency of representing a given state in a particular basis and we introduce a corresponding optimization problem.
We will always assume that our adaptive Hilbert Space bases are related to the original fixed basis by means of a unitary transformation $U \in \mathcal{G}$ belonging to a connected Lie group that is locally generated by a finite dimensional Lie algebra $\mathfrak{g}$.

If our quantum state in the original, fixed basis is $\kpsit$ we assume that it can be related to a \emph{reduced complexity state} $\kphit$ via
\begin{align} \label{eq:psi_U_phi_relation}
   \kpsit = U_t \kphit  \Leftrightarrow \kphit = U_t^\dagger \kpsit.
\end{align}
The generalization to mixed states is obvious, the fixed basis state $\rho_t$ and the reduced complexity state $\sigma_t$ are mutually related via
\begin{align} \label{eq:rho_U_sigma_relation}
   \rho_t = U_t \sigma_t U_t^\dagger  \Leftrightarrow \sigma_t = U_t^\dagger \rho_t U_t.
\end{align}
In the following we will work with mixed states for full generality. Special results applying only to pure states will be discussed as they arise.

By itself, equations \eqref{eq:psi_U_phi_relation} and \eqref{eq:rho_U_sigma_relation} give an over-parameterization of the original state. In the next section we outline how to remove this redundancy by deriving additional constraints on $\sigma_t.$

We will usually assume an explicit, smooth parametrization $U_t \equiv U_{\theta_t}$ by a set of coordinates $\theta_t \in D \subset \RR^n$. Explicit parametrizations allow us to obtain analytic insight into the coupled dynamics of the reduced complexity quantum state and the corresponding group element. We will sometimes drop the explicit time index $t$ and write $U, \phi, \psi, \rho, \sigma, \theta$ when it is clear that they are to be taken at the same time coordinate.
This, however, depends on our ability to derive explicitly \footnote{And in a form that can be efficiently evaluated.} the partial derivatives of the transformation with respect to the coordinates
\begin{align}\label{eq:Frt_definition}
    \Fr{j} := i U_{\theta}^\dagger \partialD{U_{\theta}}{\theta^j}.
\end{align}
Note that these \emph{right generators} $\{\Fr{j},\; j=1,2,\dots,n\}$ are necessarily Hermitian elements of the group's Lie algebra $\mathfrak{g}$.

In Appendix~\ref{sec:lie_construction} we outline how to derive the right generators for some types of parametrizations. In general, however, this requires the ability to explicitly compute exponentiated matrices of the adjoint representation of $\mathfrak{g}$. For high-dimensional Lie algebras this can be fairly challenging.
In Section \ref{sec:coordinate_free_method} we formulate a version of our method that does not require an explicit parametrization of the transformation.

Although our method allows for arbitrary unitary representations of Lie groups some group manifolds cannot be fully parametrized by a single coordinate patch. This can lead to additional technical difficulties which we will usually avoid by limiting ourselves to a single convex coordinate patch $\theta \in D \subset \RR^n$ that includes the origin $0$ which we always assume to map to the identity $U_0 = \mathbbm{1}$.
This poses no serious limitation in most cases of interest.
Finally, note that any such parametrization is not unique. We can smoothly re-parametrize the coordinates and then derive the transformed generators via the chain rule. Our use of upper indices for the coordinates and lower indices for the generators is thus motivated by their covariant and contravariant transformation under such re-parametrizations.

\subsection{The complexity functional} 
\label{sub:the_complexity_functional}

Consider first a canonical example: for a single bosonic degree of freedom with lowering operator $a$, and a transformation group given by coherent displacements $U_\theta = D(\alpha) = e^{\alpha a^\dagger - \alpha^\ast a},$ an intuitive constraint would be to demand that $\expect{a}_{\sigma_t}=0,$ or equivalently $\expect{a}_{\rho_t} = \theta.$
This fully fixes the coordinates and removes the redundancy. This is precisely the constraint which the QSD software package implements. As we demonstrate below, this approach is equivalent to finding the coherent displacement coordinates of the nearest -- in the sense of minimal quantum relative entropy distance -- symmetric variance Gaussian oscillator state.
This approach works well for nearly coherent states $\rho_t$ but as we saw in Section~\ref{sub:example_application_the_degenerate_parametric_oscillator} it can actually \emph{increase} the complexity when $\rho_t$ has significant non-Gaussian features.

Below we re-derive this as the result of an optimization problem, which enables us to generalize and improve the approach.
To formulate the problem, we introduce a complexity functional $\CF(\theta; \rho)$ which, given a state $\rho$, attains a unique global minimum on the space of coordinates $\theta$, we can then fix our coordinates at all times to be
\begin{align}
    \label{eq:CF_min}
    \theta_{t} := \theta^{(\CF)}_\ast(\rho_t) = {\rm{ argmin}}_{\theta} \; \CF(\theta; \rho_t).
\end{align}

\subsection{Expectation  minimization} 
\label{ssub:expectation_minimization}

The simplest choice of complexity functional is obtained by evaluating the expectation of a lower bounded operator $M$ that penalizes population of undesired basis levels in the transformed basis.
\begin{align}
    \CF_{M}(\theta; \rho) := \underbrace{\expect{U_\theta M U_\theta^\dagger}_{\rho}}
                                    _{\expect{M}_{U_\theta^\dagger \rho U_\theta}}.
\end{align}
For a bosonic degree of freedom, the penalty operator $M$ could simply be the canonical number operator $M = a^\dagger a$. As we will see below, when the transformation is given by coherent displacements $U_{\theta}:=D(\theta^1+i\theta^2)$ this particular choice leads to the QSD scheme $\theta_\ast = \left(\re{\expect{a}_{\rho}},\im{\expect{a}_{\rho}}\right).$

More generally, however, \emph{any} lower bounded operator defines a partial order relation for its eigenspaces via the ordering of its eigenvalues. According to such an order, a low complexity state would be characterized by being confined to a subspace spanned by basis states of low order.
For composite systems comprising multiple degrees of freedom or even just competing measures of complexity for a single degree of freedom, we can define composite penalty operators by taking linear combinations of $e$ mutually commuting, positive penalty operators $\{M_k \; | \; M_k \ge 0, [M_k,M_l]=0,\; k,l=1,2,\dots,e\}$ with positive weights $\{\beta_k \ge 0,\; k=1,2,\dots,e\}$, i.e.,
\begin{align}
    M(\beta) = \sum_{k=1}^{e} \beta^k M_k,\quad \beta^k > 0,\; k=1,2,\dots, e.
\end{align}
We discuss how to choose the weights $\beta$ below.
We will generally refer to this class of optimization problems as \emph{expectation minimization}.

\subsection{Counting Operators} 
\label{par:counting_operators}

For single degrees of freedom there often exists a special class of such positive operators that has an evenly spaced and non-degenerate spectrum $\lambda_{k} = \lambda_0 + \lambda k.$ Examples of this are the number operator for a bosonic degree of freedom and any projection to a single spatial axis of the angular momentum operator such as $J_z$ for a system with conserved total angular momentum $\vec J ^2 = \hbar^2 J(J+1).$ Any such operator can be normalized such that $\lambda_0=0$
 and $\lambda=1$ by rescaling and translation $M\to (M-\lambda_0 \mathbbm{1})/\lambda$.
We will refer to such normalized operators as \emph{single degree counting operators}
\begin{align}
    N = \sum_{k=0}^{d} k \ket{k}\bra{k},\; d \in \NN \cup \{\infty\},
\end{align}
where $\{\ket{k}; k=0,1,\dots,d\}$ is the orthonormal eigenbasis of $N$.
Such counting operators may also admit certain raising and lowering operators $A_{\pm} = A_\mp^\dagger$ satisfying $[N,A_\pm] = \pm A_{\pm}.$
This implies that $A_\pm \ket{k} \propto \ket{k\pm 1}$.
It is then possible to show that under the additional condition that $[A_-,A_+] = \alpha_1 \mathbbm{1} + \alpha_N N$ for some real numbers $\alpha_1, \alpha_N$ this reduces exactly to the above mentioned examples, i.e., in the infinite dimensional case $d=\infty$ the above conditions imply that our problem is equivalent to a harmonic oscillator with the typical raising and lowering operators $A_+\propto a^\dagger$, while the finite dimensional case $d=2J+1$ is equivalent to an angular momentum space of fixed integral or half-integral $J$ with the angular momentum ladder operators $A_\pm \propto J_\pm$.

As above we can combine individual counting operators $\{N_k, k=1,2,\dots, e\}$ to define composite counting operators
\begin{align}
    N(\beta) := \sum_{k=1}^{e} \beta^k N_k,\quad \beta^k > 0,\; k=1,2,\dots, e.
\end{align}
In most cases of interest to us the resulting operators have a unique ground state $ N(\beta) \ket{\Omega} = 0$, and they always have a very simple spectrum. An important special case is realized when the individual counting operators $N_k$ count excitations of different physical subsystems and each has its own pair of raising and lowering operators $[N_k,A_{l,\pm}]=\pm \delta_{kl}A_{l,\pm}.$
It is easy to see that these must still be raising and lowering operators of the composite counting operator
\begin{align}
    [N(\beta),A_{l,\pm}] = \pm \beta^k A_{k,\pm}.
\end{align}

In some cases there exist additional generalized raising and lowering operators for the composite counting operator beyond the raising and lowering operators associated with individual subsystems corresponding to particle exchange or correlated particle creation. 
As an example, for a collection of harmonic oscillators with raising and lowering operators $\{a_1, a_1^\dagger, \dots, a_e, a_e^\dagger\}$ we could define $N(\beta) = \sum_{k=1}^e \beta^k a^\dagger_k a_k$ as a composite counting operator and we would then find that the quadratic operators $\{a_j^\dagger a_k,\; j,k=1,2,\dots, e\}$ which induce unitary mixing of multiple oscillator modes or $\{a_j^\dagger a_k^\dagger,\; j,k=1,2,\dots, e\}$ which -- along with their hermitian conjugates -- induce multi-mode squeezing are also generalized raising and lowering operators:
\begin{align}
    \left[\sum_{k=1}^e \beta^k a^\dagger_k a_k,\; a_l^\dagger a_m\right] &= (\beta^l-\beta^m) a_l^\dagger a_m \\
    \left[\sum_{k=1}^e \beta^k a^\dagger_k a_k,\; a_l^\dagger a_m^\dagger \right] &= (\beta^l+\beta^m) a_l^\dagger a_m^\dagger \\
    \left[\sum_{k=1}^e \beta^k a^\dagger_k a_k,\; a_l a_m \right] &= -(\beta^l+\beta^m) a_l a_m.
\end{align}
These are useful for reducing representation complexity associated with multi-degree correlations.
When combining several individual counting operators, it may not be a priori known how to pick good weights $\beta := (\beta^1,\dots,\beta^e)^T$. In Section \ref{sub:quadratic_expansion_of_relative_entropy} we will see that $\rho$ itself contains all the information required to choose $\beta$.

We will generally refer to this class of optimization problems, i.e., minimizing the expectation of single or composite counting operators, as \emph{linear excitation minimization}.
The expectation value of counting operators and polynomials of counting operators can generally be expressed as a finite linear combination of operator expectations with coefficients depending on the coordinates. In some cases this allows to solve for the optimal coordinates directly and in closed form.

\subsection{Iterative complexity reduction} 
\label{ssub:iterative_complexity_reduction}
While some choices of complexity functionals lead to solutions in closed form, we must usually resort to numerical optimization. For compact Lie groups represented on finite dimensional Hilbert spaces there exist some provably powerful \emph{gradient flow} methods as described in \cite{Li2008LeastSquares,Herbrueggen2010Gradient} but the non-compact and (at least formally) infinite dimensional case that is often of interest to us is more complicated and less well understood.
Nonetheless, for specific examples of groups and parametrizations we are able to prove the convexity of some linear and even nonlinear expectation minimization schemes which allows us to employ existing schemes such as generalized Newton's method \cite{Mahony2002Geometry} for obtaining optimal coordinates.
Here we derive explicit expressions for the gradient and Hessian in terms of operator moments resulting from a second order expansion of the functional
\begin{align}
\CF(\theta+\delta \theta; \rho) & =
\CF(\theta; \rho) + \sum_{j=1}^n y_j(\theta; \rho) \delta \theta^j \\
& \qquad + \frac{1}{2}\sum_{j,k=1}^n h_{jk}(\theta; \rho) \delta \theta^j \delta \theta^k + O\left(\delta \theta^3\right) \nr
\end{align}

If the complexity functional is strictly convex then the Hessian $h(\theta; \rho) = (h_{jk})_{j,k=1}^n$ is positive definite everywhere and an appropriate variant of Newton's method can be applied to find the optimal coordinates which are implicitly defined by requiring the gradient to vanish $y_j(\theta; \rho)=0,\; j = 1,2,\dots, n.$

For the case of expectation minimization with penalty operator $M$, the explicit expressions for the gradient and Hessian are also given by simple expectation values
\begin{align}\label{eq:exmin_yj_mjk}
y_j(\theta;\rho) & = \expect{Y^{>}_j(\theta)}_{\sigmatest} \\
h_{jk}(\theta;\rho) = h_{kj}(\theta;\rho) & = \expect{H_{jk}^{>}(\theta)}_{\sigmatest}
\end{align}
where
\begin{align}
    Y^{>}_j(\theta) &:= i[M, \Fr{j}] \\ \label{eq:Hessian_ops}
    H_{jk}^{>}(\theta) &:= \left[\Fr{j}, \left[M, \Fr{k}\right]\right]  + i \left[M, \partialD{\Fr{k}}{\theta^j}\right].
\end{align}
Although it is not immediately obvious from Equation \eqref{eq:Hessian_ops}, the symmetry of the Hessian operators $H_{jk}^{>}(\theta) =H_{kj}^{>}(\theta)$ follows straightforwardly from $\frac{\partial^2 U_\theta}{\partial \theta^j \partial \theta^k} = \frac{\partial^2 U_\theta}{\partial \theta^k \partial \theta^j}$.



When $\rho_t$ evolves with time we can now either solve the minimization problem \eqref{eq:CF_min} at each time and use this to obtain the coordinates $\theta_t$ or alternatively derive explicit (stochastic) differential equations for the coordinates. While the former will allow us to adapt our scheme to arbitrary stochastic dynamics -- jump equations and diffusive dynamics -- the latter method can provide us with more insight into the dynamics and open up interesting opportunities for designing control schemes.

\subsection{Nonlinear excitation minimization} 
\label{ssub:CGF_min}

In this section we explain the importance of the level spacing of the penalty operator spectrum.
Assume that we are starting with a counting operator $N$. Since $N$ is positive, we can bound the probability of highly excited states using Markov's inequality
\begin{align}
    \PP_{\rho}\left[N_\theta>N_0\right] \le \frac{\expect{N_\theta}_{\rho}}{N_0}.
\end{align}
Unfortunately, this bound decays
only as $O\left(N_0^{-1}\right).$

A useful alternative is then to exploit Markov's extended inequality by applying a monotonically increasing map to $N \to f(N).$
In this case we must have
\begin{align} \label{eq:extended_markov}
    \PP_{\rho}\left[N_\theta \ge N_0\right] = \PP_{\rho}\left[f(N_\theta)\ge f(N_0)\right] \le \frac{\expect{f(N_\theta)}_{\rho}}{f(N_0)}.
\end{align}
If $f(n)$ increases super-linearly with $n$, then the resulting penalty operator $f(N)$ has an increasing spacing of eigenvalues versus $N$ and thus the bound on high excitations will decrease faster than $O\left(N_0^{-1}\right)$.

A very strong bound can be achieved with the exponential map $f(n) = \exp(\lambda n)$ parametrized by $\lambda > 0.$
This gives
\begin{align} \label{eq:extended_markov_exp}
    \PP_{\rho}\left[N_\theta>N_0\right]& \le \expect{e^{\lambda (N_\theta - N_0)}}_{\rho} \\
    &= e^{-\lambda N_0 + \CGF_{N}}. \label{eq:chernoff}
\end{align}
where we have naturally been lead to introduce the \emph{cumulant generating function} (CGF) of the penalty operator
\begin{align}
    \CGF_{N}(\theta; \rho,\lambda) := \ln\left[\underbrace{\expect{e^{\lambda N_\theta}}_{\rho}}
                                    _{\expect{e^{\lambda N}}_{{\sigmatest}}}\right],
\end{align}
which, for very small $0 < \lambda \ll 1$ reduces to
\begin{align}
 \CGF_{N}(\theta; \rho,\lambda) \approx \lambda\expect{N_\theta}_{\rho} + \frac{\lambda^2}{2} \Var{N_\theta}_{\rho} + O(\lambda^3),
\end{align}
showing that for small $\lambda$ this optimization problem is equivalent to directly minimizing the expectation of the penalty operator while for increasing $\lambda$ also penalizing large variance.

\begin{figure}[htbp]
    \centering
    \includegraphics[width=0.8\columnwidth]{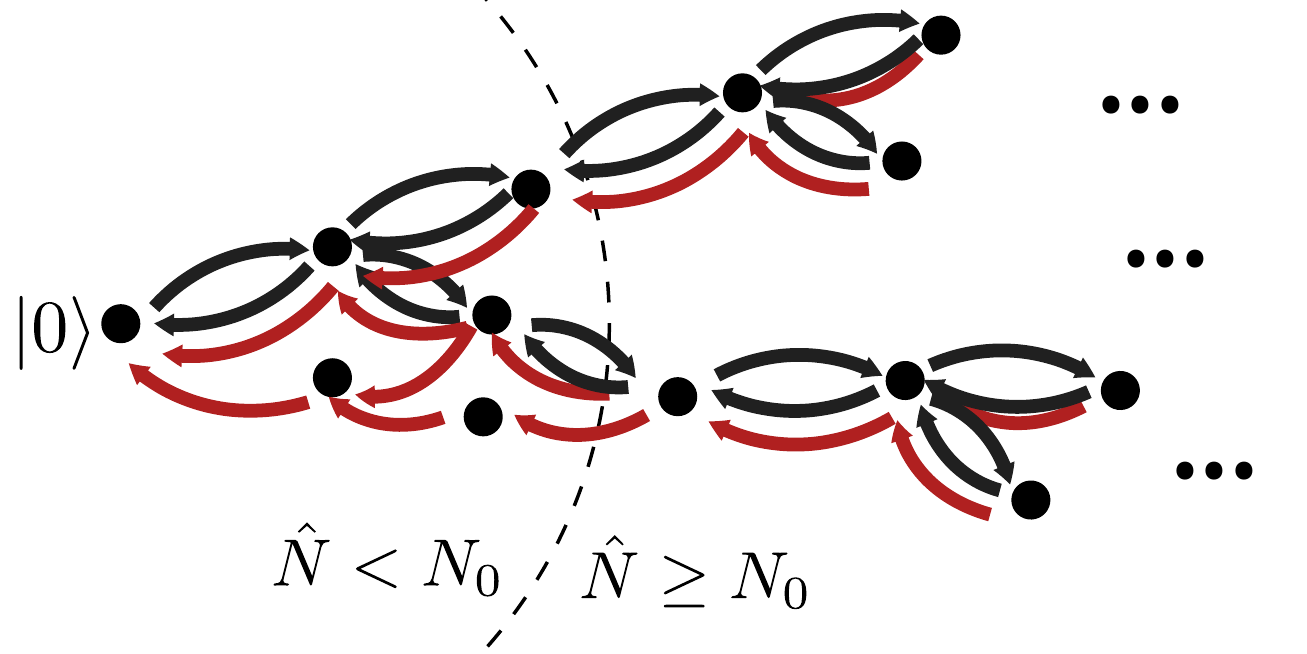}
    \caption{Many physically relevant Lie groups admit unitary representations in which there exists a natural ordering of the basis states according to the spectrum of some generalized energy or counting operator. Here we visualize such basis levels by black dots and suggest that they are (partially) ordered from left to right.
    The black arrows represent coherent transitions induced by a Hamilton operator, while the red arrows indicate dissipation induced transitions. Transforming the dynamics to a parametrized basis induces a mapping from this graph to one with potentially more transitions, e.g., under a squeezing transformation $a\to \cosh r a + \sinh r a^\dagger$ which generally increases the terms of the Hamilton operator, but if the transformation coordinates are chosen wisely, the system state can be kept close to the left side of this graph, i.e., it is trapped in a low-dimensional subspace of the overall transformed basis.
    }
    \label{fig:trapped_quantum_walk}
\end{figure}

Note that when $N$ is unbounded there may exist normalizable states ${\rho}$ for which $\expect{e^{\lambda N_\theta}}_{{\rho}}$ diverges for any $\lambda > 0,$ but this is generally true for $\expect{N_\theta}_\rho$ itself.
We will assume here without proof that such states do not actually arise in the dynamical evolution of open quantum systems.
For a constantly spaced, unbounded spectrum if there exist constants $N_0 \in \NN$ and $ \alpha \in [0,1)$ such that
 $\forall n\geq N_0$ we find
 \begin{align*}
  \frac{\bra{n+1} \sigmatest \ket{n+1}}{\bra{n} \sigmatest \ket{n}} \leq \alpha,
 \end{align*}
 then $\expect{e^{\lambda N_\theta}}_{{\rho}}$ exists for all $\lambda < \ln 1/\alpha.$
The inequality in \eqref{eq:extended_markov_exp}-\eqref{eq:chernoff} is an example of a Chernoff bound. Since the bound is satisfied for any $\lambda$ we can minimize the right hand side over $\lambda$ to achieve the most restrictive bound yielding
\begin{align}\label{eq:min_over_lambda}
    \ln \PP_{\rho}\left[N_\theta \geq N_0\right] \leq  \min_{\lambda} \CGF_{N}(\lambda)-\lambda N_0.
\end{align}
Given $N_0$ the optimal $\lambda_\ast$ leading to the lowest bound is implicitly defined via
\begin{align}\label{eq:CGF_legendre}
    \left.\partialD{\CGF_{N}(\lambda)}{\lambda}\right|_{\lambda_\ast} = N_0.
\end{align}
The negated left hand side in equation \eqref{eq:min_over_lambda} above is proportional to the digits of relative accuracy obtained when truncating the basis at the level $N_0.$ The above demonstrates that it is related to the cumulant generating function via a Legendre transformation:
\begin{align}
     -\ln \PP_{\rho}\left[N_\theta \geq N_0\right] \ge \mathcal{A}(N_0) := \sup_{\lambda} \lambda N_0 - \CGF_{N}(\lambda).
 \end{align}
We thus see that for a fixed $N_0$ our complexity functional guarantees a minimal accuracy with which a given quantum state can be represented in a $d$-dimensional subspace of the overall state space, where $d$ is the number of eigenvalues $\lambda_k$ of $N$ such that $\lambda_k < N_0$. This is visualized in Figure~\ref{fig:trapped_quantum_walk}. Conversely, assuming that $\mathcal{A}(N_0)$ is one-to-one, for any desired accuracy there exists a specific truncation level $N_0$ at which the Chernoff bound is tightest, making it the most efficient truncation level to achieve a certified accuracy.

We will refer to this family of nonlinear excitation minimization schemes as \emph{CGF minimization}.

\section{Information theoretic interpretation} 
\label{par:information_geometric_interpretation}

In this section we demonstrate that expectation minimization can also be interpreted as minimizing the quantum relative entropy between $\rho$ and a transformed canonical Gibbs state $\chi_{\theta,\beta}:=U_{\theta} \chi_\beta U_{\theta}^\dagger$ where $\chi_\beta := Z(\beta)^{-1}e^{-\beta M}$ and $Z(\beta) = \Tr{e^{-\beta M}},$ which is itself the canonical Gibbs state associated with the transformed penalty operator
\begin{align}
 \chi_{\theta,\beta} = Z(\beta)^{-1}\exp{\left(-\beta \underbrace{U_{\theta} M U_{\theta}^\dagger}_{=:M_{\theta}}\right)}\,.
 \end{align}
To see this, note that the quantum relative entropy between $\rho$ and $\chi_{\theta,\beta}$ is given by
\begin{align}
    S\left(\rho \| \chi_{\theta,\beta}\right) & = \Tr{ \rho \left[\ln\rho - \ln\chi_{\theta,\beta}\right]}  \\
     & = - H(\rho) - \expect{\ln \chi_{\theta,\beta}}_{\rho} \\
     & = - H(\rho) + \ln Z(\beta) + \beta \expect{M_{\theta}}_{\rho} \label{eq:qrel_entropy}.
\end{align}
Here the first and second terms $-H(\rho) + \ln Z(\beta)$ do not depend on $\theta$ and therefore minimizing the quantum relative entropy over all possible coordinates $\theta$ is equivalent to excitation minimization as discussed above.
We may also minimize \eqref{eq:qrel_entropy} over $\beta$ to derive an optimal Gibbs weight $\beta_\ast$ for a given quantum state $\rho$ and the associated optimal coordinates $\theta_\ast.$
By the construction of the Gibbs state, this is equivalent to $\rho$ and $\chi_{\theta, \beta_\ast}$ having equal expectation values of the penalty operator, i.e.,
\begin{align}\label{eq:expectation_equivalence}
\expect{M}_{\sigma} = \left.-\partial_\beta \ln Z(\beta)\right|_{\beta_\ast} =& \expect{M}_{\chi_{\beta_\ast}} \\
\Leftrightarrow \expect{M_{\theta_\ast}}_{\rho} =& \expect{M_{\theta_\ast}}_{\chi_{\theta_\ast,\beta_\ast}}.
\end{align}
This then affords a very nice interpretation of expectation minimization: It finds the transformation coordinates that minimize the quantum relative entropy between the actual state and a manifold of generalized Gibbs states. These form the set of maximum Von-Neumann entropy states constrained to have a specific expectation of the penalty operator. In this sense, given a penalty operator $M$, they form the \emph{the least biased manifold of states to compare a given state with}.

When $\beta=\beta_\ast$ the minimum relative entropy is given by the difference in entropies of the actual state and the nearest thermal state
\begin{align}
    S(\rho || \chi_{\theta, \beta_\ast}) & = H(\chi_{\theta_\ast, \beta_\ast}) - H(\rho) \\
        & = H(\chi_{\beta_\ast}) - H(\sigma).
\end{align}
Note that for a pure state $\rho_t = \kpsit\bpsit$ the second term vanishes, whereas the positivity of $S(\rho||\chi)\ge 0$ (with equality if and only if $\rho=\chi$) implies that $\chi_{\theta_\ast,\beta_\ast}$ really is the maximum (over all states) Von-Neumann entropy state with minimum (over all parameters) expected penalty $\expect{M_{\theta_\ast}}_{\rho} = \expect{M_{\theta_\ast}}_{\chi_{\theta_\ast,\beta_\ast}}$. By its definition, the Von-Neumann entropy only depends on the spectrum of the density matrix, which is invariant under unitary transforms. This allows us to compute it in either the original or the moving basis $H(\rho_t) = H(\sigma_t)$.

In the specific case of collective oscillator states with $M= N(\beta) = \beta^1N_1 +\dots+\beta^e N_e$, i.e., multi-mode counting operators, the minimal quantum relative entropy between a given state and a generalized Gaussian state has been employed as a measure of non-Gaussianity \cite{Genoni2008Quantifying,Genoni2010Quantifying}. If we take $U_\theta$ to allow for all Gaussianity preserving transformation (coherent displacements, unitary and more generally symplectic transformations) then excitation minimization corresponds exactly to finding the nearest Gaussian state. Often, however, we may restrict the transformation to a sub-manifold of the full set of Gaussian transformations to avoid a large decrease of the sparsity of the operators generating the moving basis dynamics.
More generally, for any composite penalty operator $M= M(\beta) = \beta^1M_1 +\dots+\beta^e M_e$ the quantum relative entropy splits into a sum
\begin{align}
    S\left(\rho \| \chi_{\theta,\beta}\right) & =
      - H(\rho) + \ln Z(\beta) + \sum_{k=1}^e \beta^k \expect{M_{k,\theta}}_{\rho},
\end{align}
where $Z(\beta) = \Tr{\exp(-\sum_{k=1}^e\beta^k M_k)}$ and $M_{k,\theta} = U_\theta M_k U_{\theta}^\dagger$.
This information geometric interpretation then affords us a method for optimally composing penalty operators: We demand that the relative entropy be minimized not only over the transformation coordinates $\theta$ but also over the complexity weights $\beta$.

A special situation arises when the penalty operators act on different degrees of freedom, in which case we can decompose the partition function
\begin{align}
      \ln Z(\beta) = \sum_{k=1}^e \ln \underbrace{{\rm Tr}_k{\left(\exp(-\beta^k M_k)\right)}}_{Z_k(\beta^k)}.
 \end{align}
Additionally, in certain cases the group transformation may be decomposable into single degree transformations
\begin{align}
    U_{\theta} = U_{1,\theta_1} \otimes U_{2,\theta_2} \otimes \cdots \otimes U_{e,\theta_e}
\end{align}
such that each factor $U_{k,\theta_k}$ is a particular unitary operator parametrized by its own disjoint set of coordinates $\theta_k=(\theta_k^1,\dots, \theta_k^{n_k})$ and acting non-trivially only on the degree of freedom labeled $k$. In this case we have
\begin{align}
    M_{k,\theta} = U_\theta M_k U_{\theta}^\dagger = U_{k,\theta_k} M_k U_{k,\theta_k}^\dagger
\end{align}
which implies that the optimization can be carried out independently for the coordinates $\theta_k$ for each $k=1,2,\dots,e$ and also prior to optimizing over the complexity weights $\beta^k.$

\subsection{Quadratic expansion of the relative entropy} 
\label{sub:quadratic_expansion_of_relative_entropy}
Any choice of parametrized quantum state family combined with a quantitative measure of distances between quantum states induces a local distance metric for the parameters $\theta$ and $\beta$. For small coordinate displacements $\delta \theta$ and $\delta \beta$ the quantum relative entropy can be expanded to second order which yields
\begin{align}
    \delta S\left(\rho \| \chi_{\theta,\beta}\right) & =
    \sum_{k=1}^e z_k \delta \beta^k
             + \sum_{j=1}^n y_{j} \delta \theta^j \\ \nr
     & \quad + \frac{1}{2}\sum_{j=1}^n\sum_{l=1}^n h_{jl} \delta \theta^j \delta \theta^l \\ \nr
     & \quad + \frac{1}{2} \sum_{k=1}^e\sum_{j=1}^n  y_{kj} \delta \beta^k  \delta \theta^j \\ \nr
     & \quad + \frac{1}{2} \sum_{k=1}^e\sum_{m=1}^e  g_{km} \delta \beta^k  \delta \beta^m
     + O(\delta^3)
\end{align}
where the coefficients are
\begin{align} \label{eq:zk_def}
    z_k & := \underbrace{Z(\beta)^{-1}\partialD{Z(\beta)}{\beta^k}}_{-\expect{M_{k,\theta}}_{\chi_{\theta,\beta}}} + \expect{M_{k,\theta}}_{\rho}, \\
    y_{kj} & := \expect{\partialD{M_{k,\theta}}{\theta^j}}_{\rho} \label{eq:ykj_def}, \\
    y_{j} & := \sum_{k=1}^e \beta^k y_{kj}, \label{eq:yk_def}\\ \label{eq:hjl_def}
    h_{jl} & := \sum_{k=1}^e \beta^k \expect{\frac{\partial^2  M_{k,\theta}}{\partial \theta^j \partial \theta^l}}_\rho,\\
    g_{km} & :=  \frac{\partial^2 \ln Z(\beta)}{\partial \beta^k \partial \beta^m} = \Cov{M_{k,\theta}}{M_{m,\theta}}_{\chi_{\theta, \beta}}. \label{eq:gkm_def}
\end{align}
Near a minimum the linear contributions vanish and in the vicinity of such minima we thus have a quadratic form in the coordinate displacements which can be interpreted as a local distance measure.
We will use this expansion later on when we derive effective dynamics for the complexity weights. When $\rho$ is itself a member of the parametrized manifold $\rho= \chi_{\theta_\ast, \beta_\ast}$, then for $(\theta, \beta) = (\theta_\ast+\delta \theta, \beta_\ast + \delta \beta)$ the linear contribution to the above expansion clearly vanishes and the quadratic coefficients are given by the Kubo-Mori quantum Fisher information metric \cite{Petz2011Introduction}.
The operator derivatives appearing in these definitions can be transformed to the moving bases where they can be expressed in terms of the transformation generators as
\begin{align}
    \expect{\partialD{M_{k,\theta}}{\theta^j}}_{\rho} & = \expect{i[M_k, \Fr{j}]}_{\sigma} \\
     \expect{\frac{\partial^2 M_{k,\theta}}{\partial \theta^j \partial \theta^l}}_\rho &= \expect{\left[\Fr{j}, \left[M_k, \Fr{l}\right]\right]}_\sigma \\ \nr
     & \quad + \expect{i \left[M_k,\partialD{\Fr{l}}{\theta^j}\right]}_\sigma.
\end{align}
The derivatives of the partition function will generally depend on the particular choice of penalty operators, but we point out that the gradient of the relative entropy with respect to $\beta$ vanishes iff
\begin{align}
 0 = Z(\beta)^{-1}\partialD{Z(\beta)}{\beta^k} + \expect{M_{k,\theta}}_{\rho}, \quad k=1,2,\dots, e.
\end{align}
As in the case of a single penalty operator these conditions are equivalent to fixing each $\beta^k$ such that the expected penalty agrees between the Gibbs state and the actual state $\rho$:
\begin{align}
  \expect{M_{k,\theta}}_{\rho} = \expect{M_{k,\theta}}_{\chi_{\theta,\beta_\ast}}, \quad k=1,2,\dots, e.
\end{align}
This then answers our previous question from Section \ref{ssub:expectation_minimization} of how to optimally choose the weights $\beta$.
The joint minimization of the quantum relative entropy over both $\beta$ and $\theta$ can also be understood as \emph{fitting} a generalized Gibbs state parametrized by $(\beta, \theta)$ to $\rho.$


\section{Dynamics in a moving basis} 
\label{sec:dynamics_in_a_moving_basis}

In this section we describe how to derive coupled equations of motion for the transformation coordinates and the low complexity quantum state such that the coordinates rapidly converge either exactly or approximately to an optimal value.

\subsection{Pure state dynamics} 
\label{sub:pure_states}

We will start by discussing the case of pure state evolution, induced by a (stochastic) Schrödinger equation
\begin{align}
    \dkpsit = -i dG_t \kpsit.
\end{align}
The generator for a closed system and deterministic dynamics is simply $dG = \hbar^{-1}H dt,$ whereas for an open system evolving according to an unnormalized stochastic Schrödinger equation (SSE) we might have
\begin{align}\label{eq:dG}
       dG^{\rm (het)}_t = \left[\hbar^{-1}H-\frac{i}{2}\sum_{k=1}^f L_k^\dagger L_k\right] dt + i\sum_{k=1}^f   dM_t^{k\ast} L_k,
\end{align}
where the $\{L_k, k=1,2,\dots, f\}$ are the Lindblad collapse operators.
This particular complex diffusive stochastic unraveling of the open system dynamics can be interpreted as a system whose output modes are all measured using heterodyne detection with perfect fidelity.
The associated complex heterodyne measurement processes are given by $dM_t^{k} = \expect{L_k}_{\psi_t}dt + dW_{k,t}$, with complex Wiener processes satisfying
\begin{align}
     \EE{\left[\Delta W_{k}(t_0,t_1) \Delta W_{l}(t_0,t_1)\right]} &= 0, \\
     \EE{\left[\Delta W_{k}^\ast(t_0,t_1) \Delta W_{l}(t_0,t_1)\right]} &= \delta_{kl} |t_1-t_0|
 \end{align}
where
\begin{align}
    W_{k}(t) &:= \int_0^t dW_{k,t'}, \\
    \Delta W_{k}(t_0,t_1) &:= W_{k}(t_1) - W_{k}(t_0) = \int_{t_0}^{t_1}dW_{k,t'}.
\end{align}
Instead of the complex valued measurement process we could also consider single quadrature homodyne measurements. This yields
\begin{align}\label{eq:dG_hom}
       dG_t^{\rm (hom)} &= \left[\hbar^{-1}H-\frac{i}{2}\sum_{k=1}^f \left(L_k^\dagger L_k + L_k^2\right)\right] dt \nr \\
       & \quad + i\sum_{k=1}^f   dM_t^{{\rm (hom)},k} L_k.
\end{align}
Here the measurement processes are real valued $dM_t^{{\rm (hom)},k} = \expect{L_k+L_k^\dagger}_{\psi_t}dt + dY_{k,t}$ with real Wiener increments satisfying
\begin{align}
     \EE{\left[\Delta Y_{k}(t_0,t_1) \Delta Y_{l}(t_0,t_1)\right]} &= \delta_{kl} |t_1-t_0|,
\end{align}
where analogously we have introduced
\begin{align}
     Y_{k}(t) &:= \int_0^t dY_{k,t'}, \\
    \Delta Y_{k}(t_0,t_1) &:= Y_{k}(t_1) - Y_{k}(t_0) = \int_{t_0}^{t_1}dY_{k,t'}.
 \end{align}
In either case, it is essential to represent the SSE in the Stratonovich picture as this enables us to carry out stochastic projections \cite{Handel2005Quantum}.
Contrary to the usual notation we will always take $X dY$ to indicate a Stratonovich stochastic differential as opposed to an Ito differential.

We could also consider situations where some output channels are measured via homodyne and some via heterodyne measurements and we can even generalize to dynamics with discrete jumps such as the quantum jump trajectories encountered when modeling direct photon detection.

As outlined above the state vector $\kpsit$ in the fixed basis is related to the reduced complexity state vector $\kphit$ via
\begin{align}
   \kpsit = U_{\theta_t} \kphit  \Leftrightarrow \kphit = U_{\theta_t}^\dagger \kpsit\,.
\end{align}
It then follows that the transformed state evolves according to a modified SSE
\begin{align}\label{eq:dphi}
    \dkphit & =-idK_{t,\theta_t}  \kphit,
\end{align} 
with effective generator
\begin{align}
\label{eq:dKtheta}
    dK_{t,\theta_t} & :=
    \underbrace{U_{\theta_t}^\dagger dG_t U_{\theta_t}}_{=:dG_{t,\theta_t}} - \sum_{j=1}^n \Frt{j} d\theta_t^j.
\end{align}
We see that the transformed state has dynamics generated not only by the transformed SSE generator $dG_{t,\theta_t}$ but also by the explicit time dependence of the unitary mapping $- \sum_{j=1}^n \Frt{j} d\theta_t^j$. Note that the coordinate dynamics will in general be stochastic. If we were working in the Ito picture, $dK_{t,\theta_t}$ would have to include terms induced by second order stochastic differentials $d\theta^jd\theta^k$ (and even $d\theta^k dM^{j\ast}$) for which the Ito table is a priori unknown. Thus, working in the Stratonovich picture is essential as it allows us to manipulate the processes with the product and chain rules of ordinary calculus.
As of yet, we have not specified $d\theta_t^j$. In Section \ref{sec:gradient_flow_dyn} we will derive a family of dynamics that constrain the moving basis state exactly or approximately to a minimum of a complexity functional as discussed in Section \ref{sec:model_reduction_vs_model_compression}.

For pure states the expectation of any hermitian observable $X=X^\dagger$ evaluated according to the moving basis state $\phi_t$ evolves as
\begin{align}\label{eq:expect_sde_pure}
d\expect{X}_{\phi_t}
    & = -2\im{\qcor{dK_{t,\theta_t}}{X}_{\phi_t}}  \\
    & = -2\im{\qcor{dG_{t,\theta_t}}{X}_{\phi_t}} \\ \nr
    & \quad - i \sum_{j=1}^n \expect{\left[\Frt{j},X\right]}_{\phi_t}d\theta^j.
\end{align}
In deriving this, it was taken into account that \eqref{eq:dphi} does not preserve the norm.
Note that when $X$ explicitly depends on time or $\theta_t$ one needs to add additional partial derivatives accordingly.
We have used Percival's notation \cite{Percival1999Localization} for the \emph{quantum correlation}
\begin{align*}
\qcor{A}{B}_{\phi_t} := \expect{A^\dagger B}_{\phi_t} - \expect{A}_{\phi_t}^\ast \expect{B}_{\phi_t},
\end{align*}
which defines a semi-definite inner product on the space of operators (cf.~Appendix~\ref{sec:app_qcorr}).

\subsection{Mixed state dynamics} 
\label{sub:mixed_state_dynamics}

Here we discuss the case of mixed quantum states and deterministic or stochastic Lindblad master equations which are necessary when some of the output channels are unobserved or some measurements have non-ideal fidelity. Note that usually imperfect measurement can be modeled by splitting the output channel on a beamsplitter with transmissivity equal to the fidelity and then observing the transmitted output with perfect fidelity and the reflected output not at all.

The generalization to stochastic master equations is straightforward. In general the fixed basis state $\rho_t$ may evolve according to
\begin{align}\label{eq:mixed_state_evol}
    d\rho_t &= -idG_t\rho_t + i \rho_t dG_t^\dagger \\ \nr
   & \quad + \sum_{j=1}^{f'} \left[c_j \rho_t c_j^\dagger - \frac{1}{2}\{c_j^\dagger c_j, \rho_t\}\right]dt
\end{align}
where $dG_t$ is defined exactly as in the pure state case and the additional Lindblad collapse operators $\{c_1,c_2,\dots, c_{f'}\}$ correspond to unobserved output channels.
The moving basis dynamics follow from a straightforward extension of the pure state case
\begin{align}
    d\sigma_t &= -idG_{t,\theta_t}\sigma_t + i \sigma_t dG_{t,\theta_t}^\dagger\\ \nr
    &\quad  + i\left[\sum_{j=1}^n \Frt{j} d\theta_t^j,\sigma_t\right]dt \\
    &\quad  + \sum_{j=1}^{f'} \left[c_{j,\theta_t} \sigma_t c_{j,\theta_t}^\dagger - \frac{1}{2}\{c_{j,\theta_t}^\dagger c_{j,\theta_t}, \sigma_t\}\right]dt 
\end{align}
where $c_{j,\theta_t}:=U_{\theta_t}^\dagger c_j U_{\theta_t}.$
In the mixed state case, the expectations of Hermitian observables evolve as
\begin{align}\label{eq:expect_sde_mixed}
d\expect{X}_{\sigma_t}
    & = -2\im{\qcor{dG_{t,\theta_t}}{X}_{\sigma_t}} \\ \nr
    & \quad - i \sum_{j=1}^n \expect{\left[\Frt{j},X\right]}_{\sigma_t}d\theta^j \\ \nr 
    & \quad + \sum_{j=1}^{f'} \re{\expect{[c_{j,\theta_t}^\dagger, X]c_{j,\theta_t}}_{\sigma_t}}dt.
\end{align}
Here we have overloaded the notation for the quantum correlation \cite{Percival1999Localization} for mixed states in the natural way, i.e.,
\begin{align*}
\qcor{A}{B}_{\sigma_t} := \expect{A^\dagger B}_{\sigma_t} - \expect{A}_{\sigma_t}^\ast \expect{B}_{\sigma_t},
\end{align*}
We thus find almost the same result as for pure states except for an additional contribution $\re{\expect{[c_{j,\theta_t}^\dagger, X]c_{j,\theta_t}}_{\sigma_t}}$ for each unobserved output channel.

\subsection{Gradient flow coupled dynamics}
\label{sec:gradient_flow_dyn}

Assume now that we have fixed a complexity functional with Hessian $(h_{jk}(\theta_t))$ and gradient $(y_j(\theta_t))$ and that at a given time $t$ we are starting at optimal coordinates, i.e., we have already solved for $\theta_t$ such that the complexity gradient $y_j(\theta_t) = 0,\; j=1,2,\dots,n.$
Then we can determine the coordinate increments $d\theta_t^j$ by requiring $y_j(\theta_{t+dt}) = y_j(\theta_{t})+ dy_j(\theta_{t})= 0$.
We may then derive the coordinate dynamics by computing the differential change of the gradient coefficients $d \yrt{j}$ as a function of $\dkpsit$ and $d\theta_t$ solve for $d\theta_t$ such that $d \yrt{j}=0$.

More generally, if we assume that we are not starting exactly at optimal coordinates but close to the optimum, then we can instead choose a decay parameter $\eta > 0$ and solve for $d\theta$ such that
\begin{align} \label{eq:sensitivity_constraint}
d \yrt{j} \stackrel{!}{=} -\eta \; \yrt{j}\; dt, \quad j=1,2,\dots, n.
\end{align}
 This reduces to the above case when we are already at the optimum coordinates, but for a good choice of $\eta$ it leads to increased robustness to slight deviations as they are exponentially damped over time.

Inserting $X=Y_j^{>}(\theta)$ into \eqref{eq:expect_sde_mixed} while accounting for its explicit dependence on $\theta$ we obtain
\begin{align}\label{eq:dy}
dy_j(\theta_t) = \sum_{k=1}^n \mrt{jk} d\theta_t^k - dq_j(\theta_t),
\end{align}
where we have defined the \emph{bias flow}
\begin{align}\label{eq:bias_flow}
    dq_j(\theta_t) & := 2\im{\qcor{dG_{t,\theta_t}}{\Yrt{j}}_{\sigma_t}} \\ \nr
    & \quad - \sum_{l=1}^{f'} \re{\expect{[c_{l,\theta_t}^\dagger, \Yrt{j}]c_{l,\theta_t}}_{\sigma_t}}dt.
\end{align}
The second contribution to the bias flow only arises for mixed state dynamics with unobserved output channels.

Combining Equations \eqref{eq:sensitivity_constraint} and \eqref{eq:bias_flow} we find for $j=1,2,\dots, n$
\begin{align}
\sum_{k=1}^n \mrt{jk} d\theta_t^k &=  dq_j - \eta \; \yrt{j}\; dt.
\end{align}
Assuming a strictly convex complexity functional and thus a positive definite Hessian $h(\theta_t) = (\mrt{jk})_{j,k=1}^n$ this relationship can be solved for the coordinate differentials
\begin{align}\label{eq:optimal_coordinate_dynamics}
d\theta_t^k &= \sum_{j=1}^n \mrtinv{kj}  \left[dq_j - \eta \; \yrt{j}\; dt\right],
\end{align}
for $k=1,2,\dots, n$.
We point out that the bias flow is linear in $dG_{t,\theta_t}$ and has a single contribution per $c_{j,\theta_t}$ which simplifies its derivation.

We will briefly review the independent contributions of Hamiltonian and dissipative deterministic contributions as well as stochastic terms in $dG_{t,\theta_t}$.
For a Hamiltonian and thus hermitian contribution $\hbar ^{-1} H dt=\hbar ^{-1} H^\dagger dt$ we find
\begin{align}
    dq_{H,j} = -i\expect{[\hbar ^{-1}H, Y_j^{>}(\theta_t)]}_{\sigma_t} dt.
\end{align}
For an anti-hermitian dissipative drift term $Vdt = \frac{1}{2i}L^\dagger L dt$ we find
\begin{align}
    dq_{V,j} & = 2\im{\qcor{\frac{1}{2i}L^\dagger L}{Y_j^{>}(\theta_t)}_{\sigma_t}} dt\\
            & = \re{\qcor{L^\dagger L}{Y_j^{>}(\theta_t)}_{\sigma_t}}dt\\
            & = \cov{L^\dagger L}{Y_j^{>}(\theta_t)}_{\sigma_t}dt.
\end{align}
Finally, each complex diffusion term $dQ = idM^\ast L$ contributes
\begin{align}
    dq_{dQ,j} & = 2\im{\qcor{idM^\ast L}{Y_j^{>}(\theta_t)}_{\sigma_t}} \\
            & = -2\re{\qcor{dM^\ast L }{Y_j^{>}(\theta_t)}_{\sigma_t}}
\end{align}
Considered as a function of $\sigma_t$ and $\theta_t$ this can be used to derive the corresponding Ito SDEs which, however, are in general quite complicated.

\subsection{Gradient coupled fiducial state dynamics} 
\label{sub:fiducial_state_dynamics}

If the system state remains localized near a semi-classical manifold of generalized Gibbs states $U_\theta \Omega U_{\theta}^\dagger,$ then it may be advantageous to evaluate the Hessian $(\mrt{jk})$ and the bias flow $(dq_j)$ not in the actual moving basis state $\sigma_t$ but in a reference state $\Omega = Z(\beta)^{-1} e^{-\sum_{l=1}^e\beta^l M_l}$ instead. In this case, Equation \eqref{eq:optimal_coordinate_dynamics} becomes
\begin{align}\label{eq:optimal_coordinate_dynamics_fiducial}
d\theta^k &= \sum_{j=1}^n [h^{(\Omega)}(\theta_t)^{-1}]^{kj}  \left[dq_{j}^{(\Omega)} - \eta \; \yrt{j}\; dt\right],
\end{align}
for $k=1,2,\dots, n$.
The advantage of this is that $[h^{(\Omega)}(\theta_t)^{-1}]^{kj}$ and $dq_{j}^{(\Omega)}$ are purely functions of $\theta$ and other scalar model parameters. Thus, \eqref{eq:optimal_coordinate_dynamics_fiducial} can be understood as semi-classical equations of motion for the coordinates coupled to the true quantum state via the gradient $\eta \; \yrt{j}\; dt$.

\subsection{Information projected fiducial state dynamics} 
\label{sub:information_projected_fiducial_state_dynamics}

It turns out that for an expectation minimization functional the quantum state independent contributions to the dynamics $d\theta^{k,\rm sc} := \sum_{j=1}^n [m^{(\Omega)}(\theta_t)^{-1}]^{kj} dq_{j}^{(\Omega)}$ are fully equivalent to the manifold projection method proposed in \cite{Mabuchi2008Derivation} except that the metric on the tangent space is taken to be the Kubo-Mori metric \cite{Bengtsson2006Geometry} that naturally arises as a generalized quantum Fisher information metric associated with differential increases in quantum relative entropy \cite{Petz2011Introduction}. In fact, in this setting one may assume the complexity weights $\beta$ to be time dependent and derive coupled dynamics for them, as well. We discuss a special case of this in the following.

\subsection{Gibbs manifold projection} 
\label{sub:gibbs_manifold_projection}
Here we briefly sketch how to use our framework to obtain approximate low dimensional dynamics that arise when the state $\rho_t$ is constrained to the Gibbs-manifold parametrized by the group transformation coordinates $\theta_t$ and the complexity weights $\beta_t$ at all times.
For a Gibbs state $\chi_{\theta, \beta} = Z(\beta)^{-1} \exp\left(-\sum_{k=1}^e\beta^k M_{k,\theta} \right)$ both the quantum relative entropy as well as its linear variation $\sum_{k=1}^e z_k \delta \beta^k  + \sum_{k=1}^n y_k \delta \theta^k$ (cf.~Equations~\eqref{eq:zk_def} and \eqref{eq:yk_def}) with respect to $\theta$ and $\beta$ vanish because it is already on the Gibbs manifold we project to. We can now apply the full procedure outlined in \cite{Mabuchi2008Derivation} to also project the stochastic dynamics of $\rho_t$ onto the manifold. Generally this amounts to choosing $d\theta$ and $d\beta$ such that $dy = (dy_1, \dots, dy_n)$ and $(dz_1,\dots, dz_e)$ are orthogonal to the tangent space spanned by $d\theta$ and $d\beta$ with respect to the Kubo-Mori associated with the second order variation of the quantum relative entropy.

In the interest of brevity we simply present the result of that derivation. The coordinate increments $d\theta$ and $d\beta$  are the implicit solutions of the following linear system
\begin{align}
\sum_{l=1}^n h_{jl} d\theta_t^l + \sum_{k=1}^e y_{kj} d\beta^k &=  dq_j \\
\sum_{j=1}^n y_{kj} d\theta^j + \sum_{k=1}^e g_{km} d\beta_t^m &=  dm_k
\end{align}
with $h_{jl}, y_{kj}$ and $g_{km}$ defined as in Equations \eqref{eq:ykj_def} - \eqref{eq:gkm_def} for $\rho \equiv \chi_{\theta, \beta}$. The flow vectors $dq=(dq_1,\dots, dq_n)$ and $dm=(dm_1,\dots, dm_e)$ are given by
\begin{align}\label{eq:gibbs_bias_flow}
    dq_j & := 2\im{\qcor{dG_{t,\theta_t}}{\Yrt{j}}_{\chi_{\beta_t}}} \\ \nr
    & \quad - \sum_{l=1}^{f'} \re{\expect{[c_{l,\theta_t}^\dagger, \Yrt{j}]c_{l,\theta_t}}_{\chi_{\beta_t}}}dt \\
    dm_k & := 2\im{\qcor{dG_{t,\theta_t}}{M_k}_{\chi_{\beta_t}}} \\ \nr
    & \quad - \sum_{l=1}^{f'} \re{\expect{[c_{l,\theta_t}^\dagger, M_k]c_{l,\theta_t}}_{\chi_{\beta_t}}}dt,
\end{align}
both of which can be evaluated in the moving basis via the diagonal Gibbs state $\chi_\beta$ for which all off-diagonal operator moments conveniently vanish.
In many cases of interest the above expressions can be evaluated analytically as all operator moments become explicit functions of only $\beta$ and $\theta$, yielding coupled, low dimensional ODEs or SDEs that approximately describe the original quantum dynamics. 
We note that the transformation coordinates $\theta$ can often be directly associated with classical or semi-classical quantities such as generalized canonical position and momentum variables, but the complexity weights $\beta$ have a purely statistical interpretation as they encapsulate the uncertainty originally encoded in the exact quantum state.


\section{Examples of Manifolds} 
\label{sec:examples_manifold}

Here we present some examples of transformation groups, penalty operators and the corresponding generators and sensitivity variables.
A given transformation $U_\theta$ is characterized by its differential form $U_{\theta_t}^\dagger dU_{\theta_t} = -i \sum_{j=1}^n \Frt{j} d\theta_t^j$, its adjoint action on its own Lie-Algebra $U_{\theta_t}^\dagger XU_{\theta_t}$ and the sensitivity operators whose expectations form the gradient and Hessian of a particular expectation minimization problem. The sensitivity operators can be derived for any given complexity functional.

\subsection{Coherent displacement} 
\label{sub:coherent_displacement}

The simplest example is that of coherent displacements $\theta=(Q,P)$. Working in this representation we have
\begin{align}
    U_{\theta} &= e^{-iQp+iPq}, &
     U_{\theta}^\dagger a U_{\theta} & = a + \frac{Q + iP}{\sqrt{2}},
\end{align}
where the generators and right generators are given by
\begin{align}
     q &=\frac{a+a^\dagger}{\sqrt{2}},&
                p &=\frac{a-a^\dagger}{\sqrt{2}i}, \\
    \Frt{1} &= p+P/2,&
    \Frt{2} &= -q-Q/2.
\end{align}
Coherent displacements of different modes commute and therefore the extension to multiple oscillator modes follows trivially.

\subsubsection{Linear excitation minimization} 
\label{ssub:linear_excitation_minimization}

For the canonical penalty operator given simply by the photon number operator $M=N:=a^\dagger a$ the gradient and Hessian operators are
\begin{align}
    \Yrt{1} &= -q,\quad
    \Yrt{2} =  -p \\
    \Mrt{} &= \begin{pmatrix} 1 &  0 \\
                                  0 & 1\end{pmatrix}.
\end{align}
We see that the Hessian is constant and strictly positive definite. Since the domain of the coordinates is $\RR^2$ and thus clearly convex and closed, we always have a unique optimum.

\subsubsection{Nonlinear excitation minimization} 
\label{ssub:nonlinear_excitation_minimization}

Assuming a nonlinearly transformed penalty operator $M=f(N)$ the gradient operators are
\begin{align}
    \Yrt{1} &= i[f(N),p]
         =\frac{[a^\dagger, f(N)]-[a,f(N)]}{\sqrt{2}}\\
        & =-\frac{a^\dagger \Delta_f(N) +\Delta_f(N)a}{\sqrt{2}},\\
    \Yrt{2} & =  i[q,f(N)]=-i\frac{a^\dagger \Delta_f(N)-\Delta_f(N)a}{\sqrt{2}}
\end{align}
where we have introduced
\begin{align}
    \Delta_f(N) := f(N+1)-f(N),
\end{align}
which allows us to write down two useful (and equivalent) rules
\begin{align}
   [a^\dagger, f(N)] = - a^\dagger \Delta_f(N) \Leftrightarrow [a, f(N)] = \Delta_f(N) a.
\end{align}
The Hessian operators are then given by
\begin{align}
    \Mrt{11} &=  \Delta_f(N) + a^\dagger \Delta^{(2)}_f(N) a \\ \nr
    &\quad + \frac{a^{\dagger 2} \Delta^{(2)}_f(N) + \Delta^{(2)}_f(N) a^2}{2}, \\
    \Mrt{12} & = \Mrt{21} =
     -\frac{a^{\dagger 2} \Delta^{(2)}_f(N) - \Delta^{(2)}_f(N) a^2}{2i}, \\
    \Mrt{22} &=  \Delta_f(N) + a^\dagger \Delta^{(2)}_f(N) a \\ \nr
    &\quad - \frac{a^{\dagger 2} \Delta^{(2)}_f(N) + \Delta^{(2)}_f(N) a^2}{2}.
\end{align}
Here the second order differences have been introduced:
\begin{align}
    \Delta^{(2)}_f(N) := \Delta_f(N+1)-\Delta_f(N) = \Delta_{\Delta_f}(N).
\end{align}
To analyze under what conditions the Hessian may be positive, observe first that the Hessian operator matrix $H^{>} := (\Mrt{jk})$ factors as
\begin{align}
    H^{>} = V^\dagger \tilde H^{>} V,
\end{align}
where we have defined
\begin{align}
    \tilde H^{>} &:= \begin{pmatrix} \Gamma & \Sigma \\ \Sigma^\dagger & \Gamma \end{pmatrix},\\
    V &:= \frac{1}{\sqrt{2}}\begin{pmatrix}1 & i \\ 1 & -i \end{pmatrix} = (V^\dagger)^{-1}
\end{align}
via
\begin{align}
    \Gamma &:= \Delta_f(N) + a^\dagger \Delta^{(2)}_f(N) a \\
    \Sigma & := \Delta^{(2)}_f(N) a^2.
\end{align}
$V$ may be interpreted as a linear transformation from a real valued quadrature representation of the displacement $\delta \theta = (\delta Q, \delta P)^T$ to complex amplitudes $(\alpha, \alpha^\ast)^T = V\delta\theta.$
Since $V$ contains only scalars, the matrices $h, \tilde h$ of element-wise expectations, i.e.,
\begin{align}
     h_{jk} = \expect{\Mrt{jk}}_\sigma,\quad
     \tilde h_{jk} = \expect{\tilde{H}^{>}_{jk}(\theta_t)}_\sigma,
 \end{align}
 are related by a similarity transform $h = V^\dagger \tilde h V$ and therefore have the same eigenvalues.
 Equipped with the above definitions we are thus able to formulate and prove the following\\
\textbf{Theorem:}\label{par:theorem_f_n} A sufficient condition for the positive semi-definiteness of the Hessian $h$ is that the following operators are non-negative
\begin{align}
     0 \le & \; \Delta^{(2)}_f(N-1)\\
     0 \le & \; \Gamma - \Xi\\
     0 \le & \; \Gamma
 \end{align}
 where $\Gamma$ is defined as above and
 \begin{align}
    \Xi := \frac{a^\dagger\left[\Delta_f^{(2)}(N+1)+\Delta_f^{(2)}(N-1)\right]a + \Delta_f^{(2)}(N)}{2}.
\end{align}
 \emph{Proof:} See Appendix \ref{sec:proof_of_theorem_ref}.

This theorem is helpful because the operators appearing in its conditions are all diagonal in the eigenbasis of $N$, which implies that we can verify the positivity simply by evaluating the inequalities for each eigenstate of $N$.
Note furthermore that the non-negativity of $\Gamma$ is also a necessary condition and that when $\Gamma-\Xi>0$ is strictly positive, the Hessian is strictly positive.

Sometimes we only wish to determine whether the Hessian is positive when the states $\sigma = \Pi \sigma \Pi$ are restricted to a subspace of the full state space characterized by a projection operator $\Pi$. In this case the operator inequalities must only hold for the projected operators $\Pi\Sigma \Pi, \Gamma \Pi ,  \Xi \Pi.$

Applying the above theorem to some interesting test cases we find that $f(n) = n^2$ leads to a convex objective, as this leads to $\Delta_f^{(2)}(N-1)\ge 0, \Gamma=1+4N, \Xi = 1+2n.$ 

For $f(n) = e^{\lambda n}$ with $\lambda>0$, which is relevant for CGF minimization, we find that $\Gamma-\Xi$ is not positive everywhere, but for small enough $\lambda$ there exists an $n_\ast = \left.\left\lceil (3 x^2 - x^3)/(-1 + x)^3 \right\rceil\right|_{x=e^\lambda}$ such that at least on the subspace of number states lower than $n_\ast$ the condition holds: $\Pi_{N \le n_\ast}(\Gamma-\Xi) \ge 0.$ This upper bound on $N$ becomes arbitrarily large as $\lambda$ goes to zero.

\subsection{Squeezing and displacement} 
\label{sub:squeezed_displaced_state}
Our next example will be a mixture of squeezing and displacement albeit each only parametrized by a single variable $\theta = (Q, R)$, which does not allow to realize the most general pure Gaussian state.
\begin{align}
    U_{\theta} &= e^{-iQp}e^{iR s},& U_{\theta}^\dagger a U_{\theta} &= \frac{Q}{\sqrt{2}} + \cosh R a -\sinh R a^\dagger
\end{align}
with generators
\begin{align}
     p &=\frac{a-a^\dagger}{\sqrt{2}i}, &
                s &=\frac{a^2-a^{\dagger2}}{2i}, \\
    \Frt{1} &= e^{R}p,&
    \Frt{2} &= -s.
\end{align}
For the canonical complexity functional $N=a^\dagger a$ the sensitivity operators are
\begin{align}
    \Yrt{1} &= -e^{R}q, \quad
    \Yrt{2} = 2r=a^2+a^{\dagger 2} \\
    \left(\Mrt{jk}\right) &= \begin{pmatrix}  e^{2R} & -e^{R}q \\
    -e^{R}q &  2+4a^\dagger a
    \end{pmatrix}.
\end{align}
This still leads to a positive definite Hessian, but it is now dependent on the coordinates and the state.
It is straightforward to derive the more general case of arbitrary displacements and squeezing, but the expressions become more tedious. They are implemented in our software package QMANIFOLD \cite{Tezak2016QMANIFOLD}.

\subsection{Spin coherent displacement} 
\label{sub:spin_coherent_states}
Our final example here is for angular momentum states and a single irreducible representation labeled by $J$ such that $\mathbf{J}^2=J(J+1)$ and the spectrum of $J_z$ is given by $-J,-J+1,\dots, J-1, J.$
The commutator relationships are $[J_z,J_\pm] = \pm J_\pm$ and $[J_+,J_-]= 2J_z.$
The ladder operators can also be represented in terms of the x-y generators $J_\pm = J_x\pm i J_y.$
We can then define a unitary Lie group explicitly via
\begin{align}
    U_{\theta} &= e^{-\mu J_+}e^{-\ln(1+|\mu|^2) J_z} e^{\mu^\ast J_-},\\
    U_{\theta}^\dagger J_- U_{\theta} &= \frac{J_- - \mu^2 J_+ -2\mu J_z}{1+|\mu|^2},\\
    U_{\theta}^\dagger J_z U_{\theta} &= \frac{\mu^\ast J_- + \mu J_+ +(1-|\mu|^2) J_z}{1+|\mu|^2} \\
    \text{where } \mu &=\theta^1+i\theta^2\\
    \Frt{1} &= \frac{-iJ_- +i J_+ -2\theta^2 J_z}{1+|\mu|^2}\\
    \Frt{2} &= \frac{-J_- -  J_+ +2\theta^1 J_z}{1+|\mu|^2}.
\end{align}
The action of this unitary on a fixed reference state, e.g., $U_\theta \ket{j_z=-J}$ can be identified with the spin coherent states as introduced by Radcliffe \cite{Radcliffe2001Some}.
For the canonical spin complexity functional $N=J_z$ the sensitivity operators are
\begin{align}
    \Yrt{1} &=  \frac{-J_- - J_+}{1+|\mu|^2},
    \quad \Yrt{2} =   \frac{iJ_- -i J_+}{1+|\mu|^2}\\
    \left(\Mrt{jk}\right) &= { \frac{4}{1+|\mu|^2} \begin{pmatrix}  \re{\mu J_-} -J_z &  \im{\mu J_-} \\
        \im{\mu J_-} & -  \re{\mu J_-} -J_z \end{pmatrix}},
\end{align}
where $\re{\mu J_-} = (\mu J_- + \mu^\ast J_+)/2$ and $\im{\mu J_-} = (\mu J_- - \mu^\ast J_+)/2i$.
This does not generally lead to a positive definite Hessian, e.g., consider the $j_z=+J$ eigenstate of $J_z$ for which the expected Hessian is clearly negative definite.
For any given state the eigenvalues of $\left(\expect{\Mrt{jk}}_\sigma\right)$ are given by
\begin{align}
    \lambda_\pm = \frac{4}{1+|\mu|^2} \left[-\expect{J_z}_\sigma \pm |\mu| \sqrt{\expect{J_x}^2_\sigma + \expect{J_y}^2_\sigma}\right]
\end{align}
Assuming that $\expect{J_z}_\sigma < 0$, these are non-negative iff
\begin{align}
    |\mu|^2 \left[\expect{J_x}^2_\sigma + \expect{J_y}^2_\sigma\right] \leq \expect{J_z}^2_\sigma.
\end{align}
Two sufficient but not necessary conditions for this are given by
\begin{align}
    |\mu|^2 < \frac{\expect{J_z}_\sigma^2}{J(J+1)-\expect{J_z^2}_\sigma},
\end{align}
and a more restrictive version that depends only on $\expect{J_z}_\sigma^2.$
\begin{align}
    |\mu|^2 < \frac{\expect{J_z}_\sigma^2}{J(J+1)-\expect{J_z}^2_\sigma}.
\end{align}

This follows directly from $\expect{J_k}_\sigma^2 \le \expect{J_k^2}_\sigma$ for $k=x,y,z$ and from $\sum_{k\in\{x,y,z\}} J_k^2 = \mathbf{J}^2 = J(J+1).$

\section{Coordinate free method} 
\label{sec:coordinate_free_method}

The methods we have introduced in the preceding section are quite appealing in that they can allow us to simulate significantly larger open quantum systems than is possible with static bases. They can also yield analytic insight into semi-classical dynamics and provide more intuition. With the analytic results presented in Section \ref{sec:examples_manifold} one can construct commuting product transformations for fairly complex systems, but they are unlikely to work well when considering non-factoring group manifolds of coordinate dimensions beyond $O(10)$ as it becomes very difficult to obtain a fully exponentiated parametrized transform from which we can derive the right generators $\{\Frt{j}\}$. These, however, are essential to the derivation of most other important quantities.
Moreover, having an explicit coordinate based representation of the unitary $U_\theta$ is useful for transforming states from the static to the moving basis. There are, however, examples of groups that can at least in principle be employed without ever having to derive an explicit parametrization of $U_\theta$. Here we sketch out how this can be achieved in principle.

Consider a Lie group and a state space and dynamics for which
\begin{enumerate}
	\item the model operators that generate the dynamics $H,L$ and $c$ can be represented as polynomials of elements of a group's finite dimensional Lie algebra, and
	\item a full Hilbert space basis can be generated by repeated action of some raising operators $\{A_{k,+}\}$ that are inside the Lie algebra starting from a unique reference state $\ket{\Omega},$ which is itself fully characterized as the unique zero-eigenvalue eigenstate of an operator $M\ket{\Omega} = 0$, where $M$ is also either inside the group's Lie algebra or a polynomial of Lie algebra elements.
\end{enumerate}
In this case we may use the adjoint representation of the group transform itself as the parametrization, i.e., if $\{Y_1,Y_2, \dots, Y_q\} \subset \mathfrak{g}$ is a hermitian basis of the Lie algebra, then if a group element $g$ is unitarily represented on the Hilbert space as $U(g)$ there also exists a corresponding element in the adjoint representation of the group $S(g)$ such that $U(g)Y_k U(g)^\dagger = \sum_{j=1}^q S_{k}^j(g) Y_j.$
This adjoint representation will typically require only a polynomial number of parameters in the system size. In some cases where a part of the Lie algebra can be expressed as polynomials of some other basis elements further reduction is possible, e.g., consider two bosonic modes $a, b$ and a Lie algebra that also contains $a^\dagger b$ or $ab$. In this case, if we know how $a$ and $b$ transform under $U$ then we also know how $a^\dagger b$ or $ab$ transform.

If we start at $t=0$ with the identity $g=e$ and $S_k^j(e) = \delta_k^j,$ then at each time step we can consider arbitrary elements from the Lie algebra and write $U(g)^\dagger \delta U(g) = -i\sum_{j=1}^q \delta \mu^j Y_j.$
This implies that
\begin{align}
    \delta [U(g)Y_k U(g)^\dagger] & = U(g) U(g)^\dagger \delta U(g) Y_k U(g)^\dagger \\ \nr
    &\quad +U(g) Y_k \delta U(g)^\dagger U(g)U(g)^\dagger \\
    & =  -i\sum_{j=1}^q \delta \mu^j U(g) [Y_j, Y_k] U(g)^\dagger \\
    & = -i\sum_{j,l=1}^q \delta \mu^j c_{jk}^l U(g)Y_l U(g)^\dagger \\
    & = -i\sum_{j,l,m=1}^q \delta \mu^j c_{jk}^l S_l^m(g) Y_m \\
    & =  \sum_{m=1}^q \delta S_k^m(g) Y_m.
\end{align}
Here we have used the convention for the structure constants given in Equation \ref{eq:structure_constants} and the final two lines allow us to read off the differential change of the adjoint transformation
\begin{align}\label{eq:delta_S_adjoint}
    \delta S_k^m(g) = -i\sum_{j,l=1}^q \delta \mu^j c_{jk}^l S_l^m(g).
\end{align}
Armed with this, we can now reformulate the complexity reduction problem. Assume that we wish to find a transfom $U_\ast$
such that it minimizes a function $\mathcal{J}_M(U) = \expect{U M U^\dagger }_\rho$, i.e.,
\begin{align}
    U_\ast = {\rm argmin}_U \mathcal{J}_M(U).
\end{align}
Expanding this to second order in a a perturbation $\delta U = -i\delta \mu^j Y_j$ we find
\begin{align}
    \mathcal{J}_M(U + \delta U) &= \expect{U M U^\dagger }_\rho + \sum_{j=1}^q \underbrace{\expect{U \frac{[Y_j, M]}{i} U^\dagger }_\rho}_{y_j} \delta \mu^j \\ \nr
    & \quad + \frac{1}{2}\sum_{j,k=1}^q \underbrace{\expect{U [Y_{\{k},[M, Y_{j\}}]] U^\dagger }_\rho}_{h_{kj}} \delta \mu^j \delta \mu^k  \\ \nr
    & \quad + O(\delta \mu^3).
\end{align}
Here we have used the notation $[Y_{\{k},[M, Y_{j\}}]]$ to indicate that the expression is symmetrized over $k$ and $j$, i.e.,
$[Y_{\{k},[M, Y_{j\}}]] \equiv \frac{1}{2}\left\{ [Y_{k},[M, Y_{j}]] + [Y_{j},[M, Y_{k}]] \right\}$.
As above, the gradient and Hessian coefficients can be more efficiently evaluated in the current lowest complexity state $\sigma = U^\dagger \rho U.$
Assuming that $(h_{jk})$ is positive semi-definite at all steps, we can use Newton's method or a similar technique to identify good coordinate updates $\delta \mu$ and update our adjoint representation $S \to S+\delta S $ as specified in \eqref{eq:delta_S_adjoint} as well as the transformed state $\sigma=U^\dagger \rho U$ according to
\begin{align}
    \sigma \to \sigma + i \sum_{j=1}^q[Y_j,\sigma]\delta \mu.
\end{align}
As above, we may also combine this complexity reduction with dynamics of $\rho$ or $\sigma$, respectively. The results of our previous sections carry over with little modification.
Although the lack of a fixed parametrization of $U$ may cause us to worry about what the specific meaning of $\sigma$ really is, our initial assumptions guarantee that the adjoint representation matrix $S$ uniquely fixes how our Hamiltonian and all other relevant operators are transformed to the low complexity basis and furthermore guarantees that a representation in terms of the original basis states can at least in principle be recovered in the new basis by transforming $M_U:=U M U^\dagger$ using $S$ and then solving for the unique eigenvector $\ket{\Omega_U}=U\ket{\Omega}$ with zero eigenvalue of $M_U.$
All other basis states can then be recovered by acting on $\ket{\Omega_U}$ with the transformed raising operators $\{UA_{k,+} U^\dagger\}$. This prescription should work straightforwardly for more complicated groups such as the $SU(n)$ symmetry groups of multiple bosonic or fermionic modes or the symplectic group $Sp(2n)$ for an ensemble of oscillators or distinguishable particles.

\section{Summary of our methods}
\label{sec:method_summary}
In this work we have presented analytical and computational methods that allow to simulate closed and open quantum systems assuming deterministic -- unitary or ensemble averaged -- or stochastic dynamics affecting either pure or mixed states. 

Given a particular dynamical quantum system to be simulated, we suggest the following steps to apply our methods.
We assume as the most general case of mixed states and stochastic evolution as described in Equation \eqref{sub:mixed_state_dynamics} defined by a generator $dG$ and a set of unobserved dissipation operators $\{c_j, j=1,2,\dots, f'\}$.
\begin{enumerate}
    \item Based on intuition or prior simulations, identify the degrees of freedom that are most likely to localize. 
    \item Define a joint penalty operator $M = \sum_{k=1}^e \beta^k M_k \ge 0$ and choose a set of transformation generators $\{X_j, j=1, 2, \dots, n\}$ that are elements of a finite dimensional Lie algebra and use the methods outlined in the Supplementary Material to obtain an exponentiated form of the parametrization $U_\theta$ and consequently the right generators $\{\Frt{j} = iU^\dagger_{\theta_t}\partialD{U_{\theta_t}}{\theta^j}, j=1,2,\dots, n' \le n\}$. This generally requires explicit exponentiation of matrices in the adjoint representation and can be done using a computer algebra system.
    In Section \ref{sec:examples_manifold} we have provided a number of examples that can be combined to handle composite systems.
    \item Derive the explicitly transformed generators $dK_{t,\theta_t}$ as defined in Equation \eqref{eq:dKtheta} and the transformed dissipation operators for the unobserved channels $c_{j,\theta_t}=U_{\theta_t}^\dagger c_j U_{\theta_t}.$ This can most easily be done by writing these operators as polynomials of the transformation generators and then applying the explicitly obtained adjoint representation to transform them.
    \item Simulate the joint equations of motion for the reduced complexity state $\sigma_t$ and the coordinates $\theta_t$ defined as
    \begin{align}
        d\sigma_t 
     &= -idG_{t,\theta_t}\sigma_t + i \sigma_t dG_{t,\theta_t}^\dagger\\ \nr
    &\quad  + i\left[\sum_{j=1}^n \Frt{j} d\theta_t^j,\sigma_t\right]dt \\
    &\quad  + \sum_{j=1}^{f'} \left[c_{j,\theta_t} \sigma_t c_{j,\theta_t}^\dagger - \frac{1}{2}\{c_{j,\theta_t}^\dagger c_{j,\theta_t}, \sigma_t\}\right]dt  \\
        d\theta_t &=  h_t^{-1} \left[dq_t - \eta \; y_t dt\right],
    \end{align}
    where the expected gradient $y_t$ and Hessian $h_t$ are defined in Equation \eqref{eq:exmin_yj_mjk} and where the bias flow $dq_t$ has been defined in Equation \eqref{eq:bias_flow}. The gain parameter $\eta \ge 0$ can be freely chosen. A reasonable heuristic to picking a good value is to linearize the equation of motion for as a function of $\theta_t$ and choose $\eta \ge \min \left\{-\re{\lambda}, \lambda \in {\rm spec}(J)\right\}$ where $J$ is the Jacobian of that linearization.
    The special case of pure state simulations is handled equivalently.
    \item The resulting simulated trajectory for the reduced complexity state $\sigma_t$ can be used to test whether further improvements can be achieved by allowing more general transformations. On the other hand, if an inspection of the coordinate trajectory $\theta_t$ reveals that some parameters are nearly constant or more generally the coordinate trajectory is itself confined to some lower dimensional set then it may be possible to remove some degrees of freedom from the transformation via an embedding transformation $\tilde \theta \mapsto \theta$ where $\tilde \theta$ has fewer coordinates.  In general it can be useful to iterate the above steps a few times to identify a good parametrization.
\end{enumerate}

\section{Conclusion and Outlook} 
\label{sec:conclusion_and_outlook_manifold}
As stated at the outset, we have presented a family of methods that allow one to investigate the inherent complexity of quantum states by attempting to reduce the total number of variables required for their description. We provided an information theoretic interpretation of our method and several alternate prescriptions for deriving \emph{exact} coupled dynamics of the semi-classical group coordinates $\theta$ and the reduced complexity quantum state $\sigma$. Our method includes earlier work by Schack, Brun and Percival \cite{Schack1995Quantum} as a special case but provides a substantially larger, analytic framework that can also be applied in cases where their method does not work. 
Further, we have shown how our method connects to earlier model-reduction work on nonlinear projection of quantum models \cite{Mabuchi2008Derivation}.
Our approach is extremely flexible and suited for arbitrary Lie groups as long as their associated Lie algebras have finite dimension.
In practice the explicit coordinate representations will work best for low dimensional Lie-algebras, and we have also outlined how to implement a method that does not rely on an explicit coordinate parameterization.

This work has many promising future directions, some key examples are the application of our simulation scheme to quantum measurement and control problems. In particular, our method is very nicely suited to model noisy dispersive or high power qubit readout of superconducting quantum circuits \cite{Clerk2008Introduction} and it could be used for a more principled approach to the quantum feedback model for autonomous state preparation considered in \cite{Andersen2015}. It could also provide more rigorous simulations of the quantum effects in coherent optical Ising machines or coherent machine learning devices than existing techniques based on quasi-probabilities \cite{Takata2015Quantum,Maruo2016Truncated,Tezak2015Coherent}.

By completely projecting the reduced complexity state to the nearest Gibbs state, our method can also be used to derive further reduced order projected models such as the Maxwell-Bloch type model considered in \cite{Mabuchi2008Derivation} or a semi-classical coupled mode theory for nonlinear resonators similar to the Wigner method proposed in \cite{Santori2014Quantum}.
It would be very appealing to work out in more detail how our approach could improve our understanding of quantum feedback networks \cite{Gough2008Quantum,Gough2009Series} by conceptually separating both the node-systems and the interconnecting fields into quasi-classical and quantum components. A very promising framework for this avenue is given by the recent work of John Gough \cite{Gough2017NonMarkovian} on controlled flows.

It is also possible to extend the formalism to quantized fields, in particular it may be very useful to apply it to traveling wave fields inside quantum feedback networks with time delays. In this scenario the reduced complexity state for the bath modes could be modeled using Matrix Product States as in \cite{Pichler2015} or by an approximate delay model \cite{Tabak2015TD}.
Recent and exciting work by Sarovar et al.~\cite{Sarovar2017Reliability} addresses a slightly different question, i.e., that of which quantum models are robustly solvable on an analog quantum simulator, but using similar information geometric techniques as our approach. Ultimately, we expect the question of \emph{efficient} representation to be closely related to the question of \emph{robust} representation. 
Finally, we believe that it is worth investigating whether the inherent complexity of the quantum states a system evolves through can be related to its computational power. A formal framework for classical dynamical input/output systems has been introduced in \cite{Dambre2012}.

\appendix
\section{General construction of the coordinate transformation}
\label{sec:lie_construction}

In this secection we outline how to construct complex transformations and derive the right generators.
The simplest construction for the transformation is by chaining single parameter transformations
\begin{align}
    U_\eta & := V^{(1)}_{\eta^1}V^{(2)}_{\eta^2} \cdots V^{(n)}_{\eta^n},\\
    \text{where } V^{(j)}_{\eta^j} & := \exp(-\eta^j X_j),\; j=1,2,\dots, n
\end{align}
Thus far we allow for complex coordinates $\{\eta^j\}$ and arbitrary, i.e., not necessarily hermitian, generators $\{X_j\}$.
We assume that the generators $X_j$ are elements of a finite dimensional Lie algebra $\mathfrak{g}$. Given a basis $\{ Y_1,Y_2,\dots,Y_q\}\subset \mathfrak{g}$ for the Lie algebra with structure constants $c_{jk}^l$ implicitly defined via
\begin{align}\label{eq:structure_constants}
    [Y_j, Y_k] = \sum_{l=1}^q c_{jk}^l Y_l,
\end{align}
we represent each transformation generator in this basis as $X_j = \sum_{k=1}^q R_j^k Y_k.$
Using the structure constants, it is straightforward to compute the conjugation of any basis element by a single parameter transformation \cite{Puri2001Mathematical} to be
\begin{align}
V^{(j)-1}_{\eta^j} Y_k V^{(j)}_{\eta^j} &= \sum_{l=1}^q [\underbrace{\exp(A^{(j)}_{\eta^j})}_{=:S^{(j)}_{\eta^j}}]_{k}^l Y_l \\
\text{with } [A^{(j)}_{\eta^j}]_{k}^l & = \eta^j \sum_{h=1}^q R_j^h c_{hk}^l.
\end{align}
The matrices $A^{(j)}_{\eta^j}$ are typically very sparse and can be exponentiated symbolically using a tool such as Mathematica \cite{Research1988Mathematica} the SymPy package \cite{SympyDevelopmentTeam2014Sympy}.
The resulting matrices $S^{(j)}_{\eta^j} \in \CC^{n\times n}$ are elements of the adjoint representation of the transformation group and can be used to directly transform the generators.
As they each depend on only a single coordinate, they satisfy $S^{(j)}_{-\eta^j} = \left(S^{(j)}_{\eta^j}\right)^{-1}$.
With this, it is straightforward to see that
\begin{align}
d U_{\eta} & = U_\eta \sum_{j=1}^n  \Xre{j} d\eta^j \\
\text{with } \Xre{j} & := V^{(n)-1}_{\eta^n} \cdots V^{(j+1)-1}_{\eta^{j+1}}
    X_j  V^{(j+1)}_{\eta^{j+1}} \cdots V^{(n)}_{\eta^n}\\
    & = \sum_{l=1}^n\sum_{k=1}^q R_j^l [S^{(j+1)}_{\eta^{j+1}} \cdots S^{(n)}_{\eta^{n}}]_l^k
    Y_k.
\end{align}
We see that by requiring this particular differential form of $U_\eta$, i.e., with all generators on the right hand side, each generator is additionally transformed  $X_j \to \Xre{j}$ by all single parameter transformations that appear to its right.

The differential transformation can be equivalently expressed with the differential generators on the left side of $U_{\eta_t}$:
\begin{align}
     dU_{\eta}  &= \left[\sum_{j=1}^n \Xle{j}d\eta^j \right] U_{\eta}\\
     \text{with }\Xle{j} &= U_{\eta} \Xre{j} U_{\eta}^{-1} \\
                        & = V^{(1)}_{\eta^1} \cdots V^{(j-1)}_{\eta^{j-1}}
                            X_j  V^{(j-1)-1}_{\eta^{j-1}} \cdots V^{(1)-1}_{\eta^1}\\
     & = \sum_{k=1}^n [S^{(j-1)}_{-\eta^{j-1}} \cdots S^{(1)}_{-\eta^{1}}]_j^k X_k.
\end{align}
If we now assume that each generator is antihermitian $X_j = i F_j$ and if we restrict the coordinates to real values $\eta^j = \theta^j \in \RR$, then 
the resulting transformation is unitary. This is desirable because observables evaluated in the transformed frame $\bphit M' \kphit = \bpsit U^{\dagger}_{\theta_t} M U_{\theta_t} \kpsit$ are actually simply the conjugated observables $M' = U^{-1}_{\theta_t} M U_{\theta_t}.$

To transform any operator $M$ from the basis associated with $\kpsit$ to the moving basis $\kphit$ one first needs to express it exclusively in terms of functions of the generators
$M = f(X_1,X_2,\dots, X_n)$ where $f:\CC^n \to \CC$ is analytic in all variables, typically $f$ is a polynomial. We then have
\begin{align}
    M'_\theta &:= U_\theta^\dagger M U_\theta \nonumber \\
    & = f(X_1'(\theta),X_2'(\theta),\dots, X_n'(\theta)), \\
    \text{with } X_j'(\theta) & :=
    \sum_{k} [\underbrace{S^{(1)}_{\theta^1} \cdots S^{(n)}_{\theta^{n}}}_{=:S_{\theta}}]_j^k
    X_k.
\end{align}
We see that having the adjoint representation single parameter transformation matrices $\{S^{(j)}_{\theta^j},j=1,2,\dots, n\}$ allows us to do all necessary computations.
We remark that representing $M = f(X_1,X_2,\dots, X_n)$ generally does not imply a unique function $f$ as some generators may themselves be polynomials of the other generators.

Finally, note that there exist alternate ways \cite{Wilcox1967Exponential} of parameterizing groups and deriving partial derivatives that may come in useful in more complex cases. Our rules for chained single parameter transformations derived here can straightforwardly generalized to chained multi-parameter transformations.

For very complex parameterizations, analytical/symbolic methods may fail but in that case it should still be possible to work in a purely numerical representation that stores and integrates both $\theta$ and elements of the adjoint representation of $U_\theta.$

\section{Properties of the quantum correlation}
\label{sec:app_qcorr}
The quantum correlation is not a strictly positive definite inner product because the quantum self-correlation of an operator $A$ vanishes in any eigenstate
 \begin{align}\label{eq:qcor_degeneracy}
     A\kphit = \lambda_a \kphit \Leftrightarrow \qcor{A}{A}_{\phi_t}=0.
 \end{align}
 The sufficiency ``$\Rightarrow$'' of this condition is obvious, the necessity ``$\Leftarrow$'' follows from the Cauchy-Schwarz inequality for the regular Hilbert space inner product.

Restricted to hermitian operators $A^\dagger = A, B^\dagger = B$ the quantum correlation can be decomposed into its real and imaginary part as
\begin{align*}
    \qcor{A}{B}_{\phi_t} = \cov{A}{B}_{\phi_t} + i \expect{\frac{[A,B]}{2i}}_{\phi_t},
\end{align*}
with the symmetrized covariance function
\begin{align*}
    \cov{A}{B}_{\phi_t} = \frac{1}{2}\expect{\{A,B\}}_{\phi_t} - \expect{A}_{\phi_t}\expect{B}_{\phi_t}.
\end{align*}

\section{Proof of our theorem} 
\label{sec:proof_of_theorem_ref}
Since $\Gamma$ is non-negative, the trace of $\tilde h$ is non-negative: $\Tr{\tilde h} = 2 \expect{\Gamma}_\sigma \ge 0$. Due to the hermiticity of $\tilde h$ its eigenvalues must be real and it thus suffices to prove that its determinant is non-negative.
 For any non-negative operator $A\ge 0$ we must have $0 \le \expect{(a+a^\dagger)A(a+a^\dagger)}_\sigma$ for any state $\sigma$. Inserting $A=\Delta_f^{(2)}(N-1)$ -- which is positive by the first condition of our theorem -- this yields
 \begin{align}\label{eq:Delta2_ineq}
     0 \le & \; 2\re{\expect{\underbrace{\Delta^{(2)}_f(N) a^2}_{\Sigma}}_\sigma} \\ \nr
&\quad \; + \expect{\underbrace{a^\dagger\left[\Delta_f^{(2)}(N+1)+\Delta_f^{(2)}(N-1)\right]a + \Delta_f^{(2)}(N)}_{\Xi}}_\sigma.
 \end{align}
This inequality is satisfied not only for $\sigma$ but also for any unitarily rotated state $\sigma(\phi):= e^{i\phi N} \sigma e^{-i \phi N}$. The second expectation value in \eqref{eq:Delta2_ineq} is invariant under any such transformation, whereas it imparts a complex phase on the first expectation value.
\begin{align}
    \expect{\underbrace{\Delta^{(2)}_f(N) a^2}_{\Sigma}}_{\sigma(\phi)} = e^{2i\phi}\expect{\underbrace{\Delta^{(2)}_f(N) a^2}_{\Sigma}}_{\sigma}.
\end{align}
We can therefore always find a $\phi_\ast$ such that the first expectation is real valued and non-positive $2\re{\expect{\underbrace{\Delta^{(2)}_f(N) a^2}_{\Sigma}}_{\sigma(\phi_\ast)}} = -2\left|\expect{\underbrace{\Delta^{(2)}_f(N) a^2}_{\Sigma}}_{\sigma}\right|.$
We have thus shown that
\begin{align}
    \left|\expect{\Sigma}_{\sigma}\right| \le \expect{Xi}_\sigma.
\end{align}

Under the second condition to our theorem we have $\Gamma-\Xi \ge 0 $ which implies $\expect{\Gamma}_\sigma \ge \expect{\Xi}_\sigma.$
We thus know that the determinant of $\tilde h$ can be bounded from below as
\begin{align}
     {\rm Det}(\tilde h) = \expect{\Gamma}_\sigma^2 - |\expect{\Sigma}|^2 \ge \expect{\Gamma}_\sigma^2 - \expect{\Xi}^2_\sigma \ge 0.
\end{align}
This proves our claim. \qed

\section{Further results for simulating under CGF minimization} 
\label{sec:further_results_on_the_dopo_under_cgf_minimization}
\begin{figure}[htb]
    \centering
    \includegraphics[width=\columnwidth]{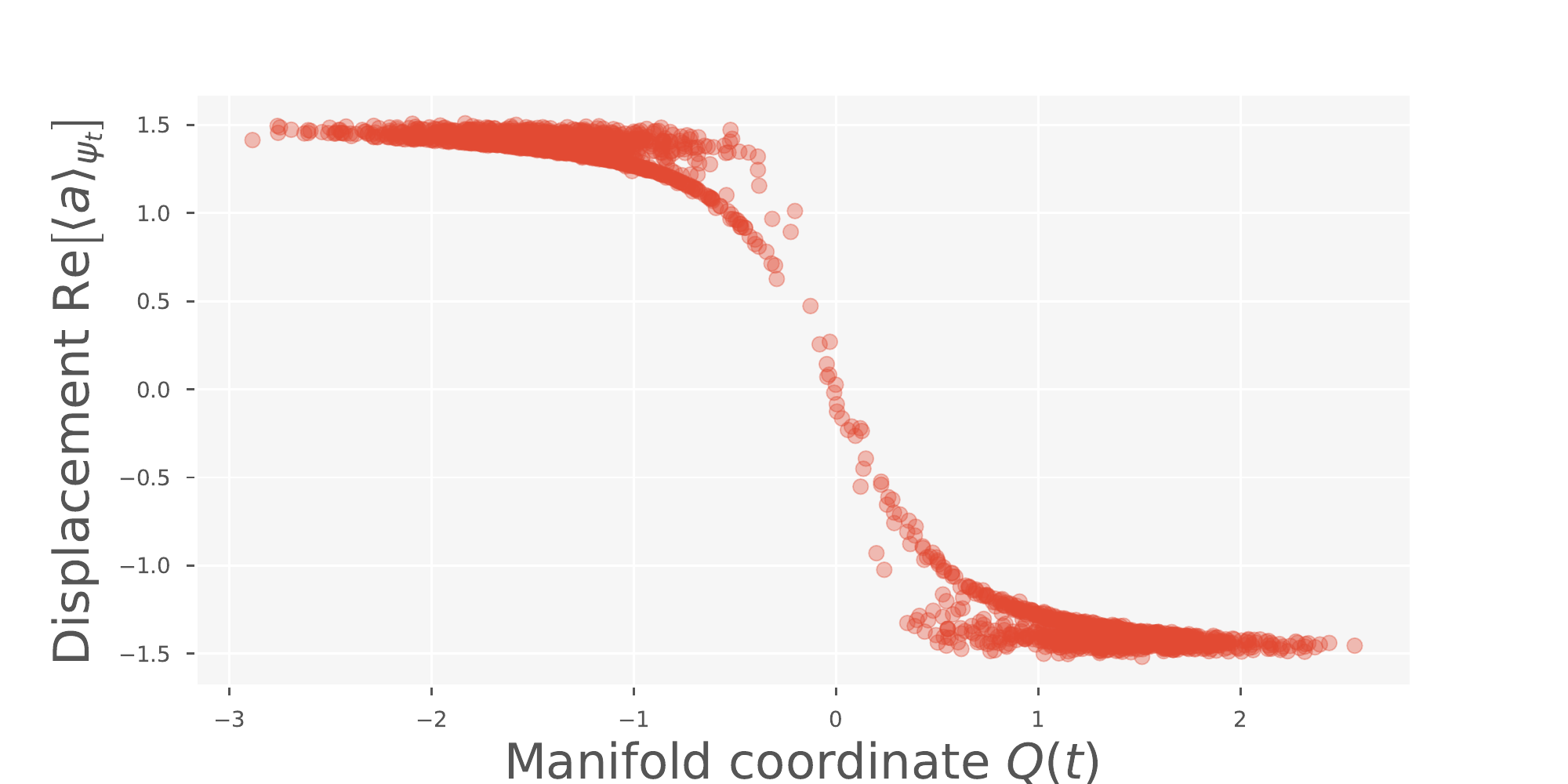}
    \caption{The optimal manifold coordinate $Q(t)$ under CGF minimization appears mostly monotonically but not linearly related to the mode expectation $\re{\expect{a}_{\psi_t}}$.}
    \label{fig:q_vs_a_exp}
\end{figure}

\begin{figure}[htb]
    \centering
    \includegraphics[width=\columnwidth]{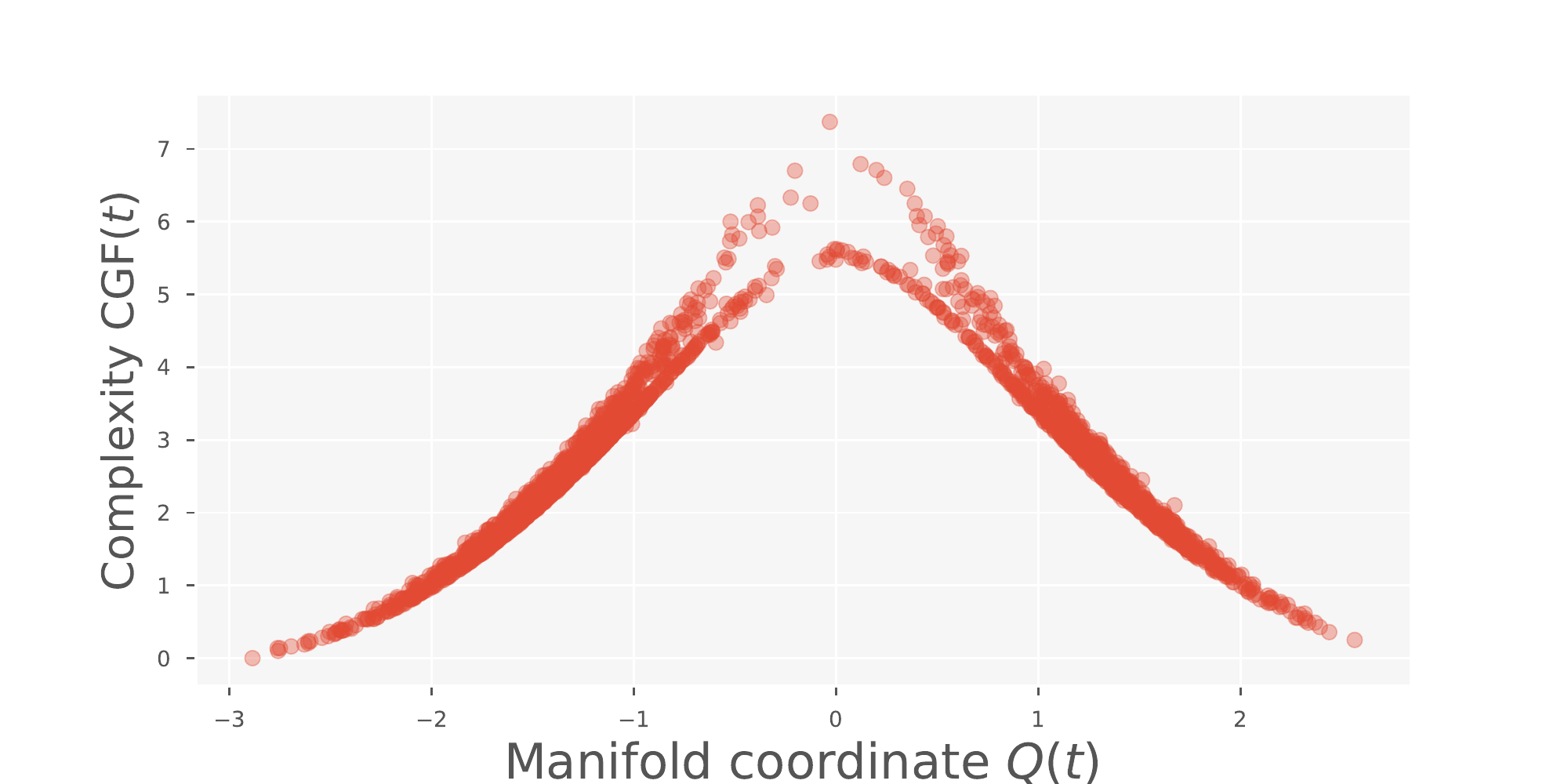}
    \caption{When the system state localizes near the `origin' i.e., $Q=0$, the complexity increases, i.e., more basis levels are necessary for accurate representation.}
    \label{fig:CGF_vs_q}
\end{figure}

\begin{figure*}[htb]
    \centering
    \includegraphics[width=.95\textwidth]{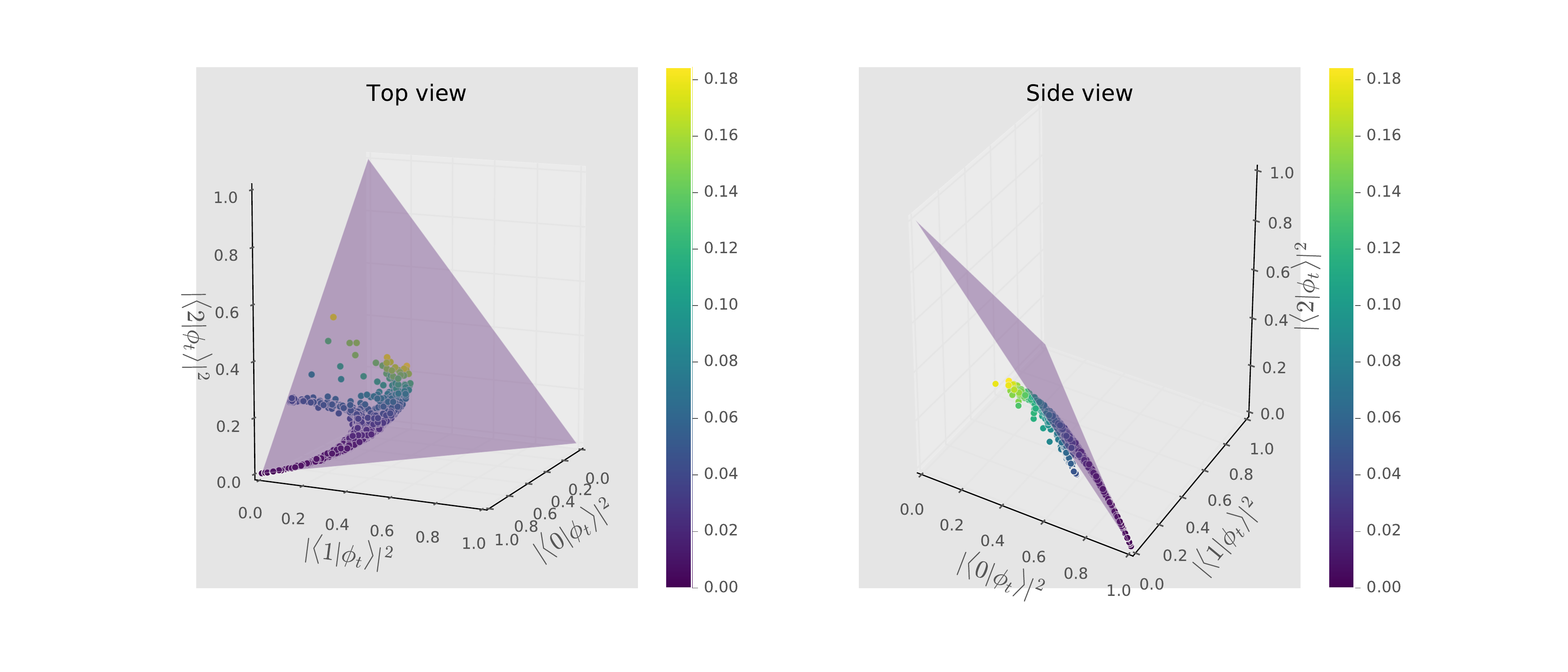}
    \caption{The probability simplex spanned by the excitation probability of the first three basis levels. For the optimal CGF trajectory the basis level populations remain nearly confined to this simplex, but diverge slightly from it, especially when the first and second excited basis levels are non-trivially populated. The points are color coded according to their `missing probability'-distance $p'$ from the simplex, i.e. $p':=1-\expect{\Pi_0+\Pi_1+\Pi_2}_{\phi_t}$.}
    \label{fig:simplex_CGF}
\end{figure*}

Here we present some additional results obtained in simulating the DOPO system in a basis that minimizes the CGF functional. In particular, we investigate the relationship between the manifold coordinates and the expectation of the oscillators's mode operators. While excitation minimization will always enforce $\expect{a}_{\phi_t}\equiv0 \Leftrightarrow \expect{a}_{\psi_t} = \alpha = \frac{Q+i P}{\sqrt{2}}$ CGF optimization generally does not lead to such a linear relationship as can be seen in Figure \ref{fig:q_vs_a_exp}.
We can also see that different regions in phase space lead to different complexity as measured by the CGF (cf Figure \ref{fig:CGF_vs_q}). This motivates using a simulation method in which even the basis size is adapted to the inherent complexity of the current dynamics.
Finally, we note that even within the displaced basis there appears to be yet lower dimensional attractors for the reduced complexity state $\kphit$. In Figure~\ref{fig:simplex_CGF} we have visualized the distribution of the first three moving basis level populations when transforming to the CGF optimal basis.
\begin{figure*}[htb]
    \centering
    \includegraphics[width=\textwidth]{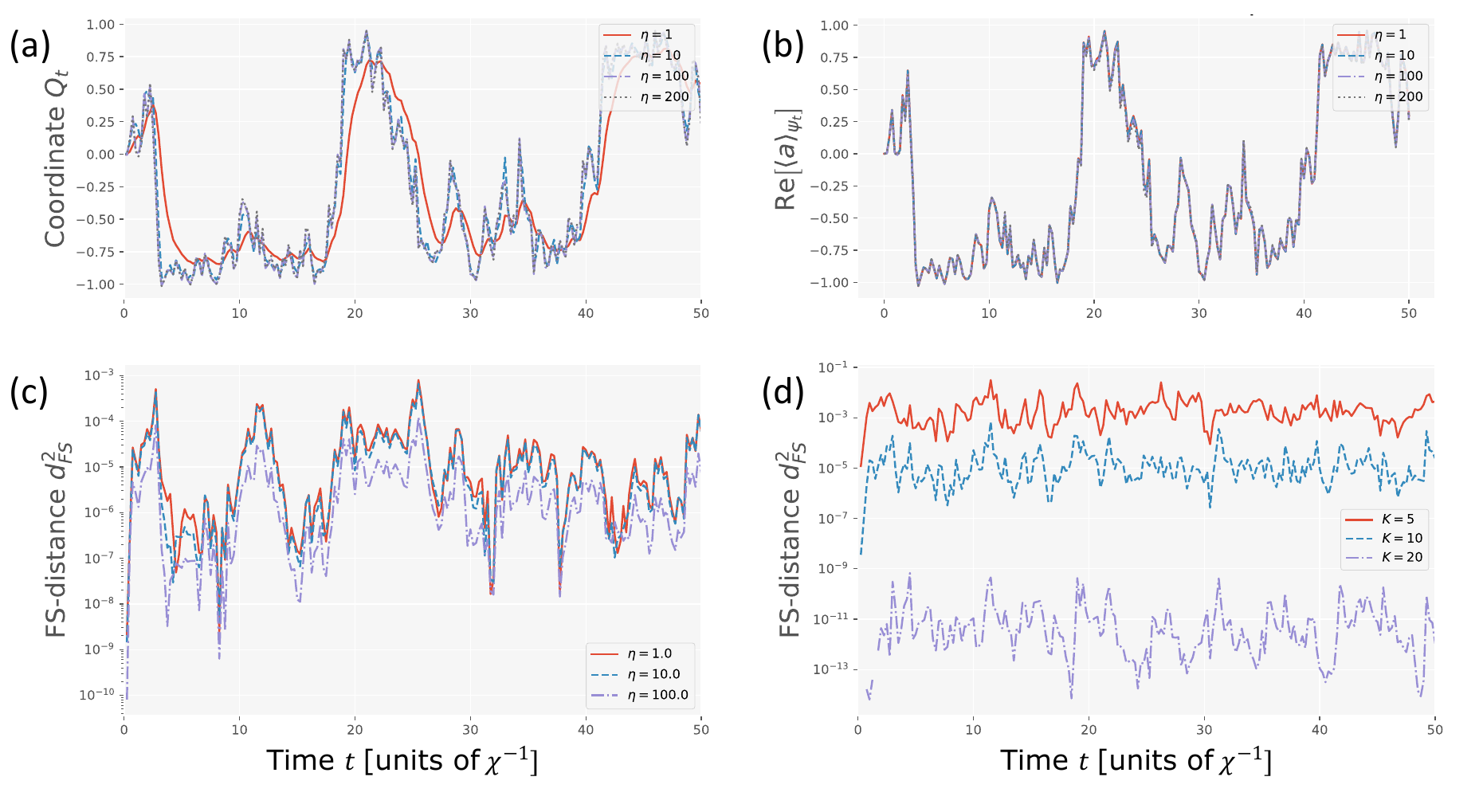}
    \caption{Gradient coupled fiducial state dynamics (simulated with $K=30$ basis levels) for varying coupling strength $\eta \in [1,10,100,200].$ In (a) we compare the $Q$ coordinate trajectories and find that they agree very well for large $\eta$ and appear somewhat `low-passed' for $\eta=1$. The mode expectation values (shown in (b)) agree very well for all values of $\eta.$ Finally, in (c) we present the Fubini-Study distance between each trajectory and the $\eta=200$ trajectory.
    In (d) we compare the Fubini-Study distance between a trajectory simulated with $K=40$ moving basis levels to trajectories carried out with $K=5,10,20$. Unsurprisingly, a larger basis decreases the error, but the errors do not appear to accumulate over time. All simulations were carried out with $\eta=10.$}
    \label{fig:gradient_coupled_dynamics}
\end{figure*}
In Figure \ref{fig:gradient_coupled_dynamics} we have simulated the gradient coupled fiducial state dynamics for the DPO system above threshold for different values of the coupling gain $\eta.$ All stochastic simulations were carried out with the same random seed and thus the same realization of the innovation process.
Figure \ref{fig:gradient_coupled_dynamics} (c) shows some discrepancies between states reconstructed from simulations with lower $\eta$ and $\eta_0=200.$
Specifically we compute $d_{\rm{FS}}^2(\psi^{(\eta)},\psi^{(\eta_0)}),$ where the fixed basis representation states are obtained from the respective moving bases representation states via $\ket{\psi^{(\eta)}(t)}=U_{\theta^{(\eta)}} \ket{\phi^{(\eta)}(t)}$ and the Fubini-Study distance is defined as:
\begin{align}
    d_{\rm{FS}}(\phi, \phi'):= \arccos{\left(|\braket{\phi}{\phi'}|\right)}
\end{align}
The errors decrease with increasing $\eta$ suggesting that a strong gradient coupling gain $\eta=O(100)$ is preferable. We intend to investigate this more systematically in future work.
We note that the errors appear to stay constant over time, which is encouraging.
Furthermore, we have simulated our system with different sizes of the basis (cf Figure \ref{fig:gradient_coupled_dynamics} (d)) and evaluated the error relative to the most accurate simulation. Surprisingly, we find that the truncation error remains roughly constant, i.e., even for this randomly switching system, the low-dimensional approximations to the system state track the actual system state very well.

\acknowledgements{
\label{sec:acknowledgements}
We wish to thank Dmitri Pavlichin, Ryan Hamerly, Tatsuhiro Onodera, Edwin Ng, Gil Tabak, Michael Goerz, Dodd Gray, Peter McMahon, Jeff Hill, Nicholas Rubin, Mohan Sarovar, Thaddeus Ladd, Hendra Nurdin, John Gough, Aashish Clerk, Patrick Hayden, Surya Ganguli, Mauro Maggioni, Todd Brun and Thomas Schulte-Herbrüggen for insightful discussion and feedback. This work was supported by the Army Research Office under grant W911NF-16-1-0086 and the National Science Foundation under grant PHY-1648807.
N.T. additionally acknowledges support through a Stanford Graduate Fellowship and a Math+X fellowship by the Simons Foundation. Also, N. H. A. is grateful for funding from the JCJC INS2I 2016 “QIGR3CF” project and the JCJC INS2I 2017 “QFCCQI” project.
}



%

\end{document}